\documentclass[conference]{IEEEtran}
\IEEEoverridecommandlockouts
\usepackage{cite}
\usepackage{amsmath,amssymb,amsfonts}
\usepackage{algorithmic}
\usepackage{graphicx}
\usepackage{subfig}
\usepackage{textcomp}
\usepackage{multirow}
\def\BibTeX{{\rm B\kern-.05em{\sc i\kern-.025em b}\kern-.08em
    T\kern-.1667em\lower.7ex\hbox{E}\kern-.125emX}}
\begin{document}

\title{An Efficient Dispatcher for Large Scale GraphProcessing on OpenCL-based FPGAs}

\author{\IEEEauthorblockN{1\textsuperscript{st} Yang Chengbo}
\IEEEauthorblockA{\textit{Huazhong University of Science and Technology} \\
Wuhan, China \\
u201214845@hust.edu.cn}
}

\maketitle

\begin{abstract}
High parallel framework has been proved to be very suitable for graph processing. There are various work to optimize the implementation in FPGAs, a pipeline parallel device. The key to make use of the parallel performance of FPGAs is to process graph data in pipeline model and take advantage of on-chip memory to realize necessary locality process. This paper proposes a modularize graph processing framework, which focus on the whole executing procedure with the extremely different degree of parallelism. The framework has three contributions. First, the combination of vertex-centric and edge-centric processing framework can been adjusting in the executing procedure to accommodate top-down algorithm and bottom-up algorithm. Second, owing to the pipeline parallel and finite on-chip memory accelerator, the novel edge-block, a block consist of edges vertex, achieve optimizing the way to utilize the on-chip memory to group the edges and stream the edges in a block to realize the stream pattern to pipeline parallel processing. Third, depending to the analysis of the block structure of nature graph and the executing characteristics during graph processing, we design a novel conversion dispatcher to change processing module, to match the corresponding exchange point. Our evaluation with four graph applications on five diverse scale graph shows that .
\end{abstract}


\section{\textbf{Introduction}}
Various objective entities can represent as graph and Graph problem has become significant fundamentality in varieties of domain including social networking, intelligent recommendation and machine learning\cite{b1}. To solve the large-scale graph problem, graph processing, a skill to execute the computation between objects and their connection, fully analyze the graph structure and acquire the relationship of the graph objects.

During the whole processing, parallelism occupies an important part owing to the large number of simultaneous vertex computing. The earliest masterpiece of graph processing mainly concentrated on the distributed system, enslaved to the plenty of communication overhead and the appropriate way to realize data partition balance. As the development of graph processing in all kinds of platform, Field-Programmable Gate Arrays (FPGAs), in particular, have become attractive because of their fine-grained pipeline parallelism, extremely low power consumption, abundant on-chip memory resources, and dedicated custom function design.

Executing large-scale graph processing in accelerators faces with some inherent difficulties including poor locality, a great deal of communication overload and irregular data access pattern, as well as new added inevitable problems including limited resources and low bandwidth size. Especially, FPGAs make the best of on-chip resources to generate the special circuits based on user¡¯s programming functions, so that generating a custom circuit requires corresponding amount of resources. Moreover, the pipeline parallelism architecture of FPGAs parallel executing task in multitasking model with streaming data pattern, resulting in data parallel processing with pipelined multitasking.

The majority of nature graphs present small-world network attribute, such as finite diameter and power law distribution characteristic\cite{b2}. Therefore, during the whole graph processing procedure, most of the vertexes and edges have been accessed or handled in a few iterations. How to utilize the pipeline parallelism of FPGA to realize parallel vertexes computation and make full use of on-chip memory and logical resources to resolve the random and frequent memory access pattern is the most dominant key to construct graph processing architecture in FPGAs. Matching the highly variable executing characteristic in graph processing and taking advantage of the on-chip resources to realize large-scale graph processing, we design the dual module processing framework. We execute the whole procedure in two different processing units from basic data structure to executing style to match the corresponding processing characteristic. We combine top-down algorithm with bottom-up algorithm to execute different parallel iteration and integrate the vertex-centric and edge-centric to match different processing procedure\cite{b3}. Basing on the factor that the majority of vertexes or edges have been accessed in a few iteration, we propose several optimization methods to make full use of on-chip resources to complete the high load execution process.

This paper makes the following contributions through the processing framework, different executing modules and evaluation of the whole architecture:

\begin{itemize}
\item \textbf{Completely different double execution module.} We execute the top-down algorithm and the bottom-up algorithm in different executing iteration to improve execution performance. In order to match the contrasting procedure and utilize the pipeline, combination of vertex-centric and edge-centric optimize the processing pattern.

\item \textbf{Edge-block executing architecture in edge-centric module.} On-chip resources in FPGA is limited and generating function units consume a certain amount circuit elements. Making full use of finite on-chip logical resource and on-chip memory to achieve executing the high parallel, we generating the block consist of eight destination vertexes to finish the high-load process.

\item \textbf{Novel executing module conversion rule.} There are various conversion strategies to exchange top-down executing style to bottom-up executing style. In our framework, we combine vertex-centric with edge-centric and adopt edge-block to optimize the on-chip memory to realize the high parallelism process.

\item \textbf{Abundant experiments about various optimization mechanism.} We devise comprehensive experiments to evaluate the framework processing performance. Furthermore, there are various optimization method in graph processing to strengthen locality or promote communication efficiency. We devise a serious of experiments to compare optimized performance of different optimization method.
\end{itemize}

The remaining of this paper is organized as follows. Section 2 introduces the background of the large-scale graph data, FPGAs parallel executing architecture and double different programming module. The whole architecture of the graph processing framework on FPGAs is shown in Section 3. The novel module conversion dispatcher and the analysis of edge-block processing is shown in Section 4. The edge-block generation rule and executing mechanism is shown in Section 5 from generation to scheduling discipline. Section 6 shows the evaluation of the processing framework and performance comparison in FPGAs with different optimization method. Section 6 discusses the related work and conclude this paper in Section 7.

\section{\textbf{Background}}
This section presents the background of graph processing framework and the pipeline and programming architecture of FPGAs. Moreover, we will introduce the comparison of vertex-centric and edge-centric programming module.

\subsection{\textbf{Graph data structure}}
Let G(V, E) represents a graph, in which V and E respectively denotes the vertex and edge sets\cite{b4}. As shown in Figure 1, the vertex in a graph represents objects and the edge represents the relationship of two vertex. Depending on the attribute of the edge direction, graph can be classified as a directed graph and undirected graph. Graph processing executes over the whole graph to gain the relationship between different vertexes or update the status or value of vertex and edge sets.

The operator of graph processing consists of three phase, including Gather, Apply and Scatter. The processing procedure of three phase is shown on the left of the Figure 2. The Gather phase gather processing information from the other objects to congregate. The Apply phase update the state with the original value and information firm the Gather phase. The Scatter phase propagates the update state to the other objects. Though the three processing operators, the graph processing utilizes several iterations to process entire graph.

\begin{figure}[tp]
\centerline{\includegraphics[width=0.38\textwidth]{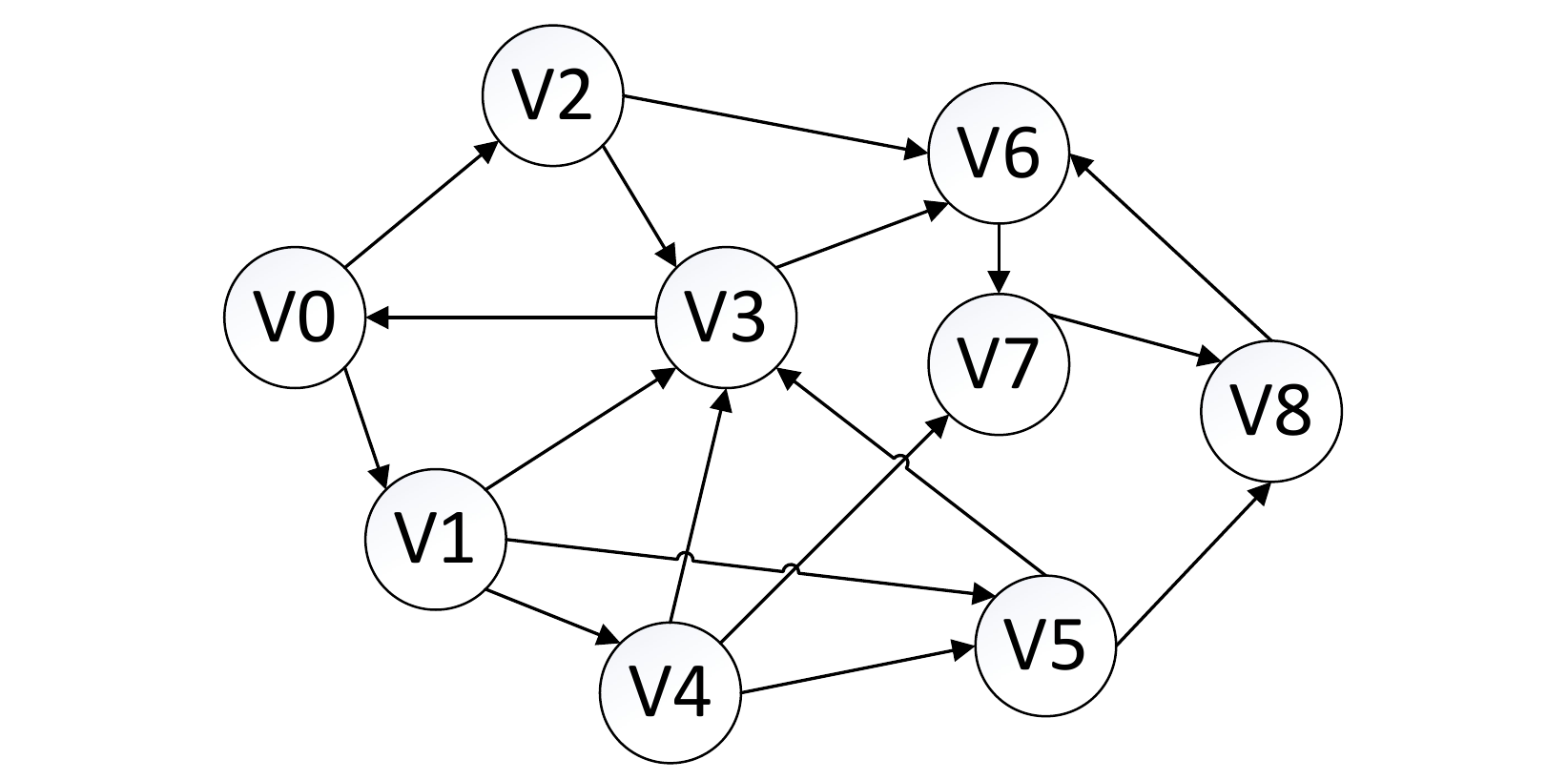}}
\caption{Graph Data}
\label{fig}
\end{figure}

\begin{figure}[tbp]
\centering
\includegraphics[width=0.17\textwidth]{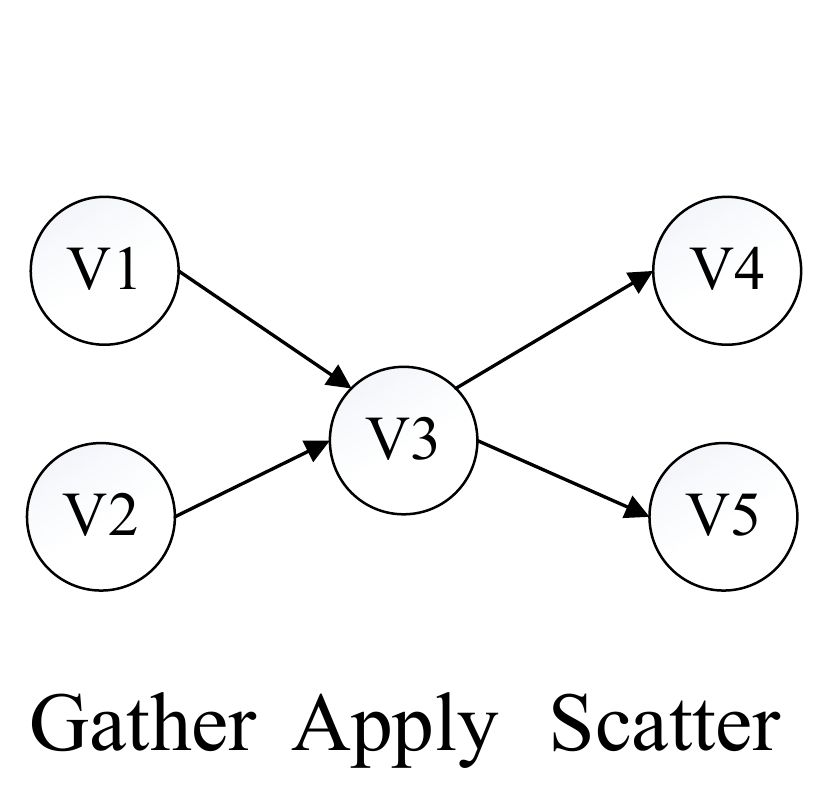}
\includegraphics[width=0.29\textwidth]{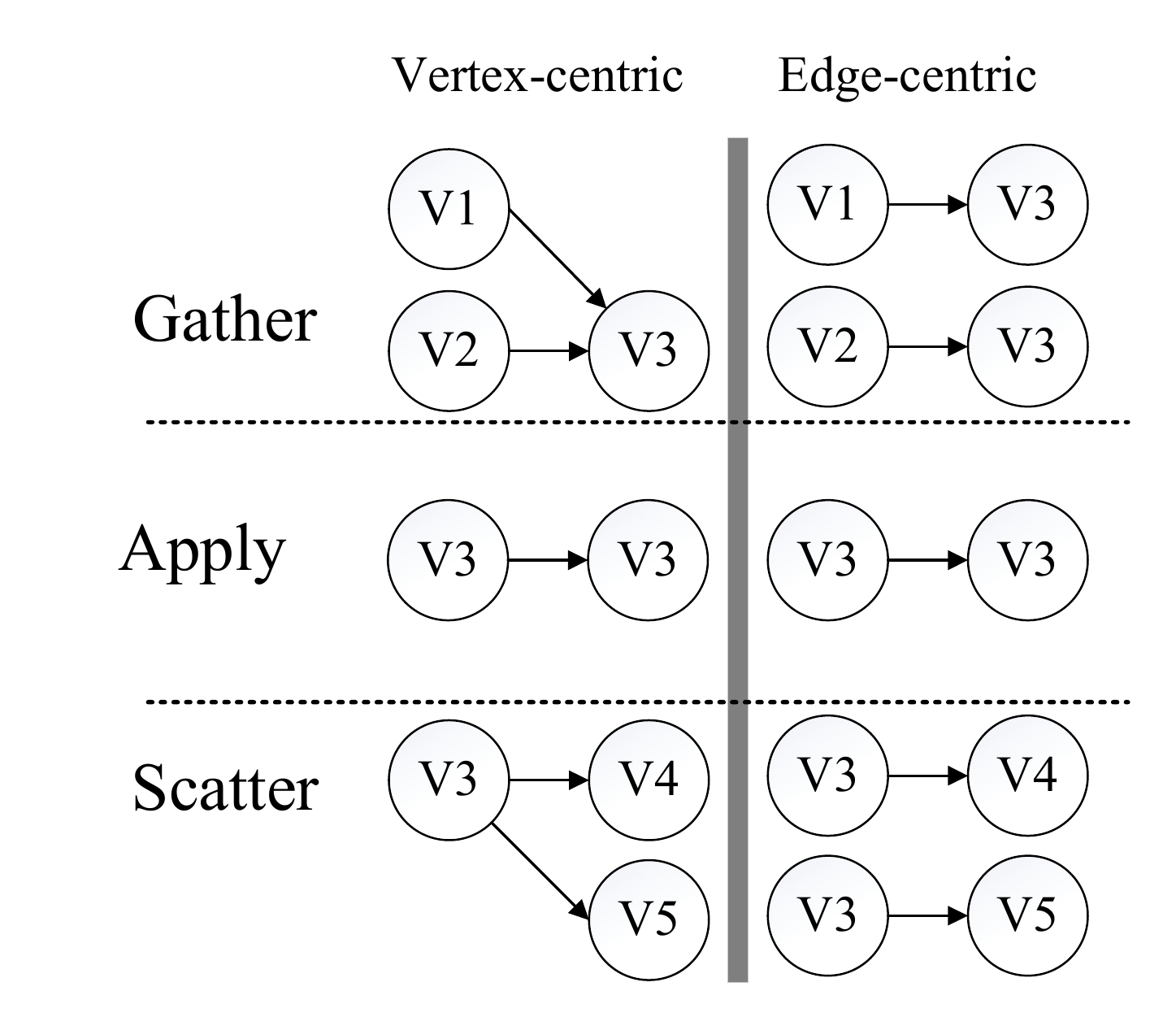}
\caption{GAS Module}
\end{figure}

\subsection{\textbf{Small-world network characteristic}}
The majority of large-scale nature graphs own small-world network characteristic: \textbf{(1) extremely short diameter compare to vertex number.} \textbf{(2) degrees of vertex set comply with power law distribution.} The diameter, quantified by the shortest distance between any two vertices in the graph, is much shorter compared to the number of vertex and edge so that vertex can be traversed from one vertex in connected subgraph among a few steps. As shown in Figure 4, the degree power law distribution results in that the majority of the vertex or edge computation concentrate on a few processing iterations and other parallel iterations only to complete a little deal of computing.

\subsection{\textbf{Compressed storage}}
The key problem to execute graph processing in accelerator is the limited memory. Apparently, to store a large-scale graph needs substantial dedicated space because of the large number of vertex and edge. However, the adjacency matrix of graph is very sparse owing to power law distribution in order to adopt compressed storage to save the storage space, such as Compressed Sparse Row (CSR) and Compressed Sparse Column (CSC).

Taking CSR as an example, CSR consists of three array. One array represents the vertex set. One array represents the edge set whose source vertex are the same in one subarray. The third array represents the index that denotes the starting position of the edge subarray whose source vertex is the index position of vertex set array. To utilize the vertex array and index array can find all destination vertex of a certain source one. If the graph¡¯s edges own weight value, CSR will add a new array whose elements correspond to edge array and store the weight value.

\subsection{\textbf{Parallelism architecture of FPGAs}}
FPGAs have been applied to various applications such as high performance computing, ¡°big-data¡±, operating system accelerator and social engine\cite{b5}. FPGAs board provides abundant resources such as logic resources, dedicated logic register and on-chip memory, as shown on the left of Figure 3. Utilizing on-chip resources properly can generating multi-threaded parallelism, but it¡¯s fundamentally diverse compare to that of CPU/GPU. FPGAs combine a range of on-chip resource to realize multitasking pipeline parallel framework and every instruction in kernel program need to take over a certain amount of resources to generating custom circuit, contrary to the general-purpose processor of CPUs/GPUs. Because of these factors, FPGAs can been much more efficient to save energy by the custom circuit, so as to been more favored in the era of energy conservation\cite{b6}.

There are several means to programing FPGAs from Hardware Description Language (HDL) to High-level synthesis (HLS). HDL can control the concrete circuit generating through hardware behavior description, structure description and data flow description. Using HDL to realize function design need to spend much time and effort to adjust programing module in order to generate underlying logic circuit with the whole and distinct comprehension to FPGAs architecture. Through decades of effort, HLS has attracted more and more interest in abstract programing ability: transform advanced programming language to custom hardware circuit to implement the program on FPGAs.

OpenCL (Open Computing Language) is a cross platform general heterogeneous programming framework, which can mine accelerators parallel processing capability entirely and collaborate with CPU to realize efficient high parallel computation. FPGAs vendors such as Altera of Intel and Xilinx have provided OpenCL SDK to support OpenCL programming on FPGAs. Altera OpenCL SDK, as a set of HLS program tools, can generate the pipeline parallelism architecture to realize concurrent data processing in massively multithreaded pattern, fully utilizing FPGAs programming framework. As shown on the right of Figure 3, Altera OpenCL SDK abstracts program away from the complicated hardware design so that system developers can concentrate on algorithm optimization and framework design. Therefore, Altera OpenCL SDK comsumes on-chip resources in FPGA to generate custom circuit for every instruction and integrates all FPGAs storage resources to form a dedicated memory model, including register, BRAM and DDRs of the FPGA board.
\begin{figure}[tp]
\centerline{\includegraphics[width=0.5\textwidth]{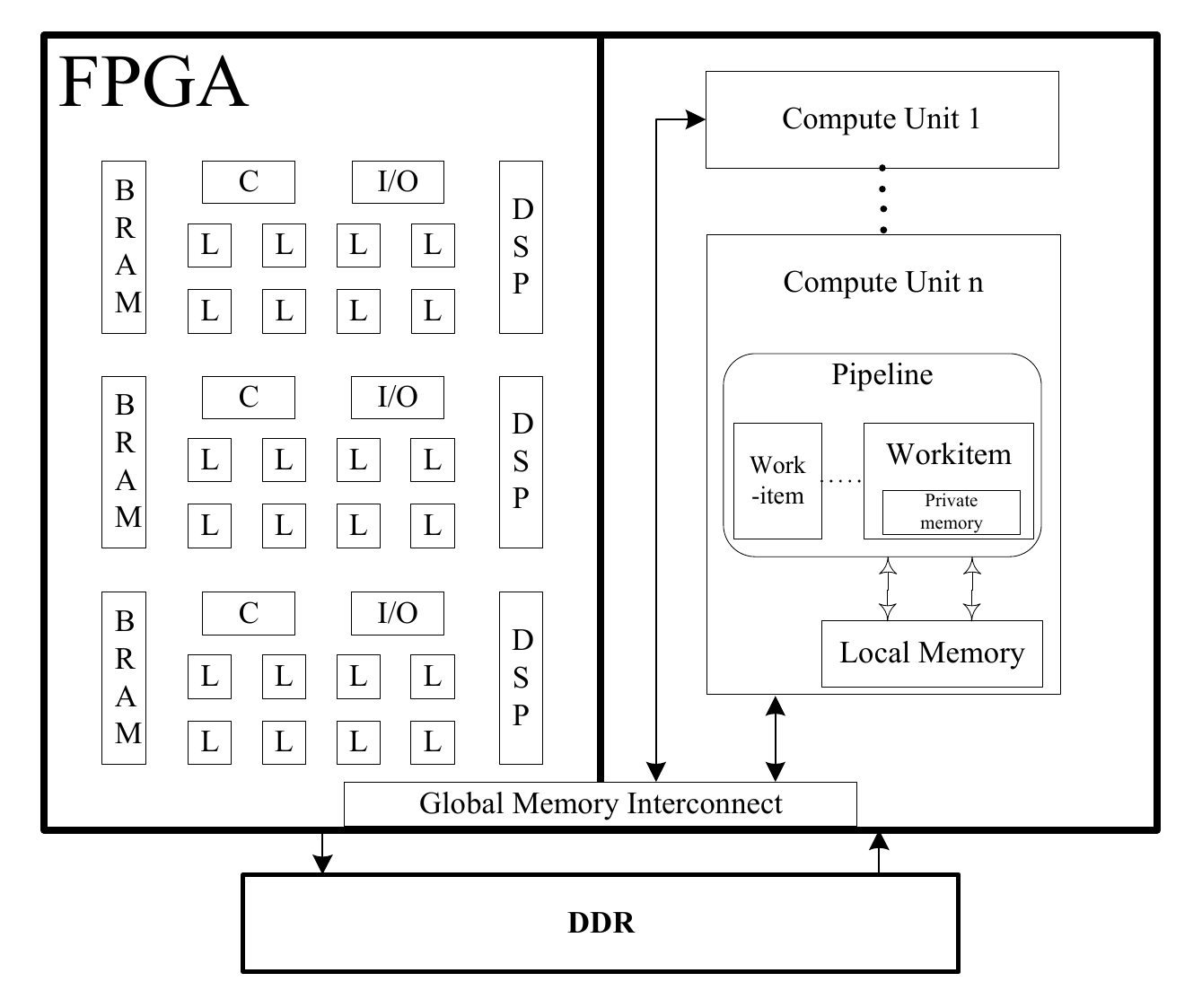}}
\caption{The Architecture of OpenCL on FPGAs}
\label{fig}
\end{figure}

\subsection{\textbf{Vertex-centric and edge-centric}}
There are two programming model to graph processing, representing two different fashion to control the graph processing iteration procedure. Figure 2 illustrates the processing procedure of the general operator on the left and shows the detail of the double model to execute three different operator phase on the right.

The vertex-centric model executes the whole processing over vertex so that edges in this model act as the role to transfer data from one vertex to another. When the data transferred from neighbors has been finished, the vertex updates its state by accumulating previous state and neighbors. After accomplishing update, the vertex propagate the state to next neighbors. The vertex centric conforms to the processing procedure of graph and has been adopted to many graph algorithms and system. Accumulating from neighbors and propagating to neighbors, the processing procedure brings about a great deal of random access and communication overload to communicate with several neighbors by means of an index to identify edges connect to a vertex.

On the other hand, the edge-centric only executes the state update on vertex set and move other operators to edge set. Before updating state, the edge set transfers the edge value to its destination vertex so that the vertex can get the data to accumulate, called the gather phase. After updating, the vertex set propagates value to edge set in order to update the edge value, called the scatter phase. Therefore, the edge-centric scatter and gather are synchronous. The edge-centric saves graph state on edge set and processes computation with consecutive edges so that all edges extracted for executing can maintain sequential access, avoiding the random fashion in vertex-centric\cite{b7}. The sequential access is very necessary and advantageous to accelerators limited to access bandwidth. The edge-centric need to represent the edge set using source and destination vertexes, so this module need more storage resources to store additional vertex set. To achieve more efficient dispatcher, we assign resources of FPGAs to schedule edge-stream based on their locality and delay\cite{b8}. However, with the accelerator memory development, accelerators can offer more storage space to store data enough.

\section{\textbf{SYSTEM ARCHITECTURE}}
This section shows the whole framework of the graph processing system. We introduce the double executing module to match the process characteristic of different phase. Furthermore, we show the details of two module to reveal the principle.

\subsection{\textbf{Modularized Processing}}
We have explain the small-world network characteristic in Section 2 and we show the process procedure owing to the finite diameter and power law distribution.

The finite diameter is very benefit for graph traversal algorithm and applications, such as breadth-first search (BFS) and single source shortest path (SSSP). The destination of graph traversal is to find all vertexes that the source vertex connect. Every vertexes can be visited among a few steps so that all vertexes in the connected subgraph can be found in a few iterations. The six degrees of separation shows the characteristic representing in real life. Therefore, the processing of graph traversal completes in a few iterations.

The power law distribution is the most important characteristic of graph processing and resulting in various causes to limit the processing performance, shown in Figure 4. The degree of a vertex is the number of edges connected and is further classified into in-degree and out-degree in the directed graph. The representation of power law distribution is that the majority of the vertex set only own a low degree and a much small number of vertex own a high degree. Moreover, there are only a few vertex own a very high degree. The factor results in the obviously different executing performance in the whole processing procedure.

\begin{figure}[tp]
\centerline{\includegraphics[width=0.46\textwidth]{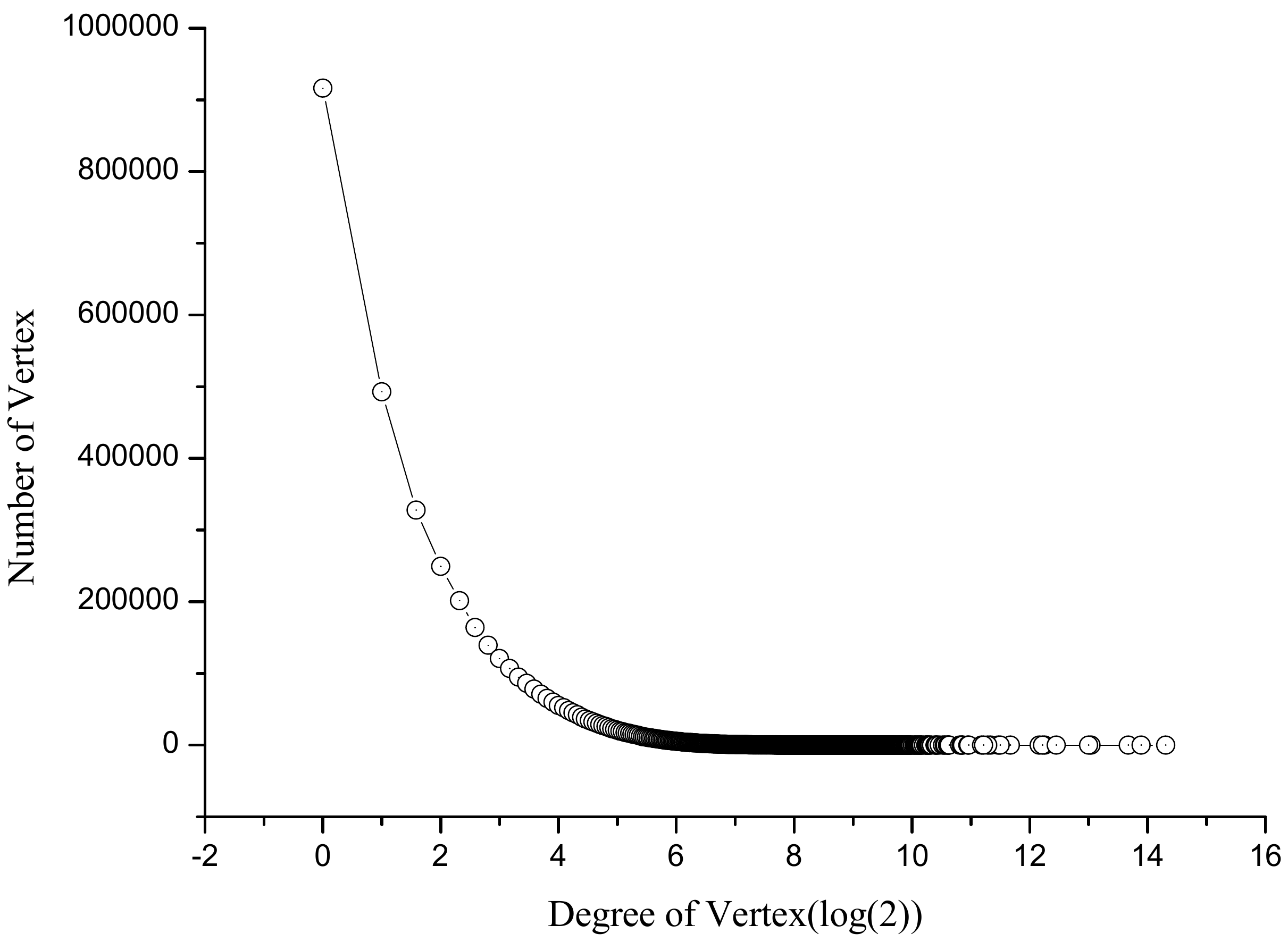}}
\caption{The Power Law Distribution}
\label{fig}
\end{figure}

Take BFS as an example to exhibit the processing procedure. BFS is to traverse vertex in levels, which all vertexes equally distant to source vertex are explored in the same level\cite{b9}. The majority of the vertex only own a low degree and very a few vertexes own a high degree so that the source vertex is much possible to be a low degree vertex. Furthermore, the neighbors of the source only owns a low degree so that the original iterations of BFS only traverse a small part of the graph. As the traversal process continuing, more and more vertexes is traversed and those ¡°hubs¡± with very high degree takes part in the frontier in every iteration very possibly, which consists of vertexes that are explored in this level and are the neighbors of next vertexes explored. Therefore, there are much more vertexes explored in these levels than original levels. In the last few iterations, the majority of vertex are explored so that the number of vertex explored in these levels become small again, the same as the original iterations. As the traversal procedure in every level can be parallel, the numbers of vertexes explored in every level represent the degree of parallelism. Therefore, the whole processing procedure of BFS has double different part with double degree of parallelism. We discuss the PAGERANK algorithm to find processing characteristic of other graph algorithms, not belonging to the graph traversal algorithm. PAGERANK is to compute the rank value of websites by counting the number and quality of links to estimate the importance of the website. In original iterations, all vertexes are active and compute the new rank value using ranking formula to accumulating the original value and neighbors¡¯. After a few iterations, the majority of vertex rank value converge and these vertexes become inactive state. Indeed, a large number of vertex become inactive after the first iteration. In the last iterations, there are a few vertexes active or active again so that only a few vertexes compute the rank value. The ranking computation can still be parallel so that the numbers of vertex ranking computation can represent the degree of parallelism the same as BFS. Therefore, the whole procedure of PAGERANK also has double part with different degree of parallelism.

\begin{figure}[tp]
\centerline{\includegraphics[width=0.4\textwidth]{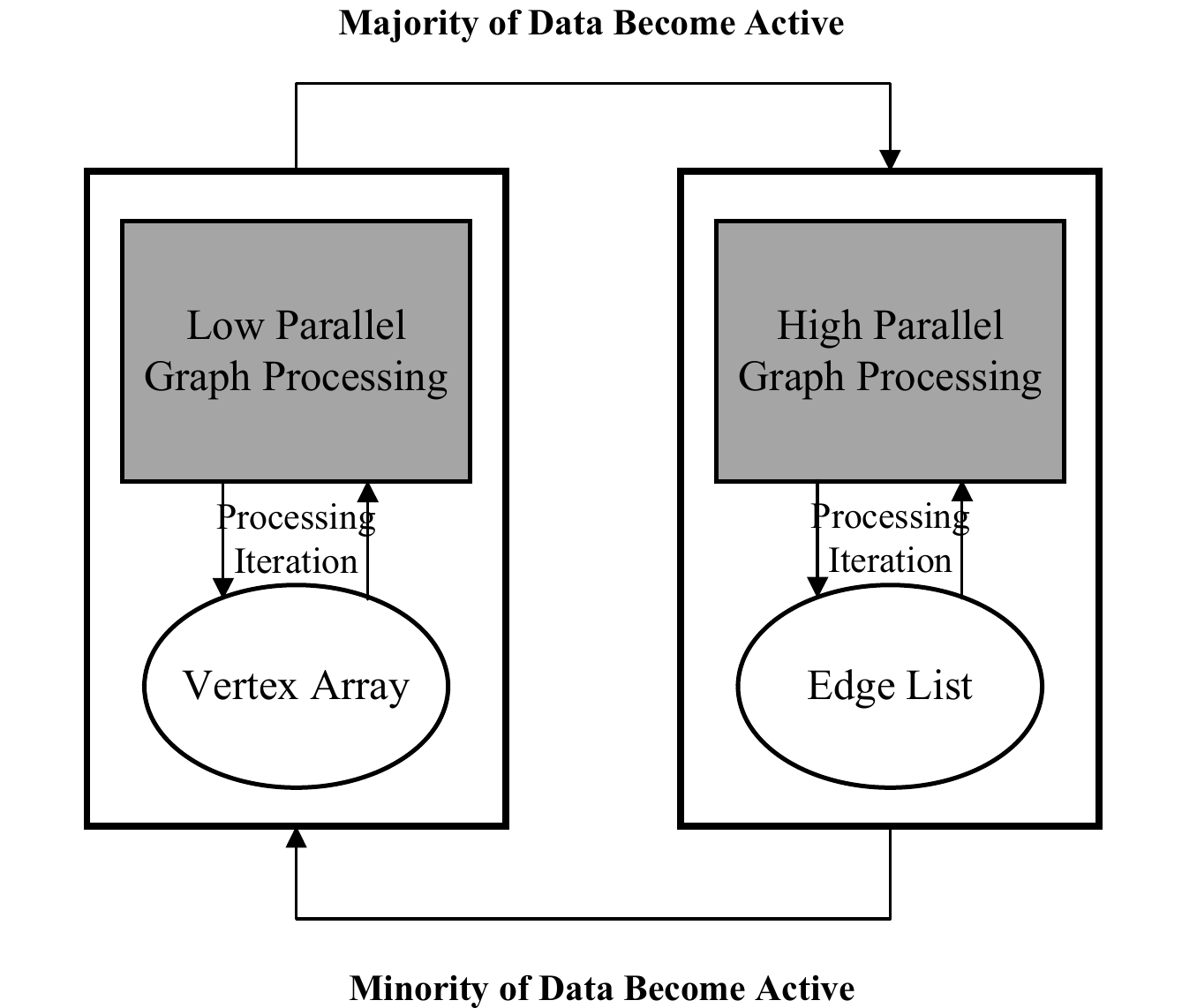}}
\caption{Modularized Processing}
\label{fig}
\end{figure}

Though discussing the processing procedure of BFS and PAGERANK, we can find that the graph processing own double executing part with different degree of parallelism. Next, we explore the connection between iteration of graph processing. The graph processing is to generate new state by treat data of the vertex set and edge set in this iteration. What the last iteration has executed doesn¡¯t influence the performance and result of this iteration so that every processing iteration is independent from others.

Basing on double kinds of degree of parallelism and independence between iterations, we design two module to process the whole procedure and match different parallel characteristic in FPGAs, as shown in Figure 5. One module is to process iterations with low degree of parallelism with vertex-centric because vertex-centric programming is suitable for communication between vertexes\cite{b10}. In the low degree parallelism stage, the number of vertex are very small and very close to each other, which almost store in the same memory block so without substantive random access. The other module is to process iterations with high degree of parallelism with edge-centric because these iterations process the most of vertexes and edges and edge-centric can realize sequential pipeline processing visualizedly. Two modules both generate in one FPGA board simultaneously so that the module conversion can finish very quickly without much time to transfer data between different processors or overload to compile another module\cite{b11}.

\subsection{\textbf{System Framework}}
The overall architecture of the graph processing mechanism is shown on the right of Figure 8. The system consists of four module on the FPGAs board. The processing system load data form DDR and store processing result to it. The logic design detail is shown in Figure 8. The processing units include a processing module for the low parallel procedure, a processing module for the high parallel procedure, a runtime dispatcher and a data analyzer.

\begin{figure}[bp]
\centerline{\includegraphics[width=0.45\textwidth]{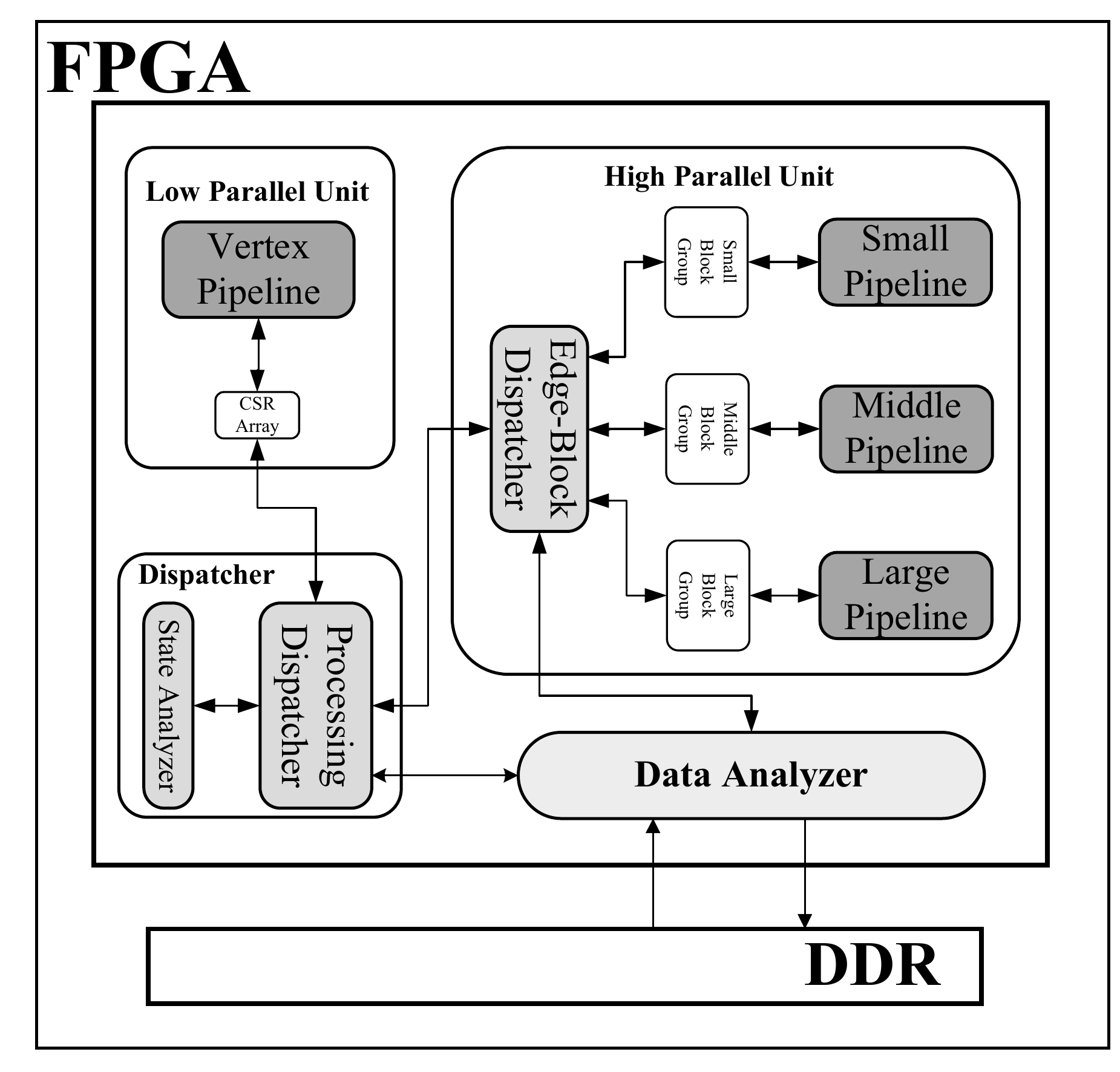}}
\caption{System Framework}
\label{fig}
\end{figure}

\begin{itemize}
\item \textbf{Data Analyzer.} The on-chip processing system needs to know the class of data loading from DDR. At the same, the system selects right data array to finish different processing procedure. The data analyzer can identify correctness to access data array from DDR.

\item \textbf{Dispatcher.} The processing system needs to monitor the runtime characteristic of processing state and realizes the transformation mechanism to match the parallelism. The Dispatcher gains the runtime characteristic to control the processing procedure and access the graph data array to dispatch data to the corresponding processing module.

\item \textbf{Low parallel Unit.} The low parallel unit is responsible for the low parallel processing section. The processing model is vertex-centric and the graph data is organized in CSR storage array. The vertex pipeline is the kernel logic for executing multi-thread pipeline to finish graph processing for every active vertex. The module contains an array cache to access the graph array. The cache gains the active vertex array from the dispatcher and sends data to the processing unit. After processing finished, the processing module store the result to the cache and the cache sends the data to the dispatcher.

\item \textbf{High parallel Unit.} The high parallel unit is responsible for the high parallel processing section. The processing model is edge-centric and the graph data is organized in edge-block array. The module contains three processing pipeline kernel logic to finish graph processing of different scale of edge-block groups. Different processing pipelines owns respective array cache to access data and the edge-block cache stores the block array ready to process or the processing result from corresponding unit. Therefore, the module contains a dispatcher to dispatch different scale of edge-block to match corresponding processing unit. The edge-block dispatcher access block array from the dispatcher module and collect different edge-block groups to generate entire graph data from different array caches.
\end{itemize}

\subsection{\textbf{Processing Mode}}
There are several existing works realizing random access decrease or communication styles optimization to optimize the performance of graph processing. The hybrid processing algorithm combines the top-down algorithm with the bottom-up algorithm is one of the most efficient method to realize great performance optimization. The top-down algorithm, called push-style, generates an active status array composed of active vertexes and the active vertex transfer its state to its neighbors. The bottom-up algorithm, called pull-style, regards the inactive vertex as active to generate the status array and the vertex fetches state of its neighbors to accumulate. In graph traversal algorithm, these traversed vertexes won¡¯t take part in processing so that the status array won¡¯t consist of these traversed vertexes.

The system integrates the hybrid algorithm to optimize the graph processing in double modules. In the module to process low degree parallelism, the number of processing vertexes is very small so that we adopt the push-style to execute the processing procedure. As shown in Figure 7, the push-style can generate the active state array easily with a few active vertexes in vertex-centric programming, which owns a reasonable locality because of short distance. In the module to process high degree parallelism, we solve the processing in edge-centric so that we update state in edge streaming, as shown in Figure 8. We need to stream the edge set to update the state as the pull-style to process the majority of data of graph because original iterations before pull-style mode only process a very small number of data. However, the mode processed in the system is different from the pure push-style because of the edge-block which we will introduce the detail in Section 5.

\begin{figure}[tp]
\centerline{\includegraphics[width=0.53\textwidth]{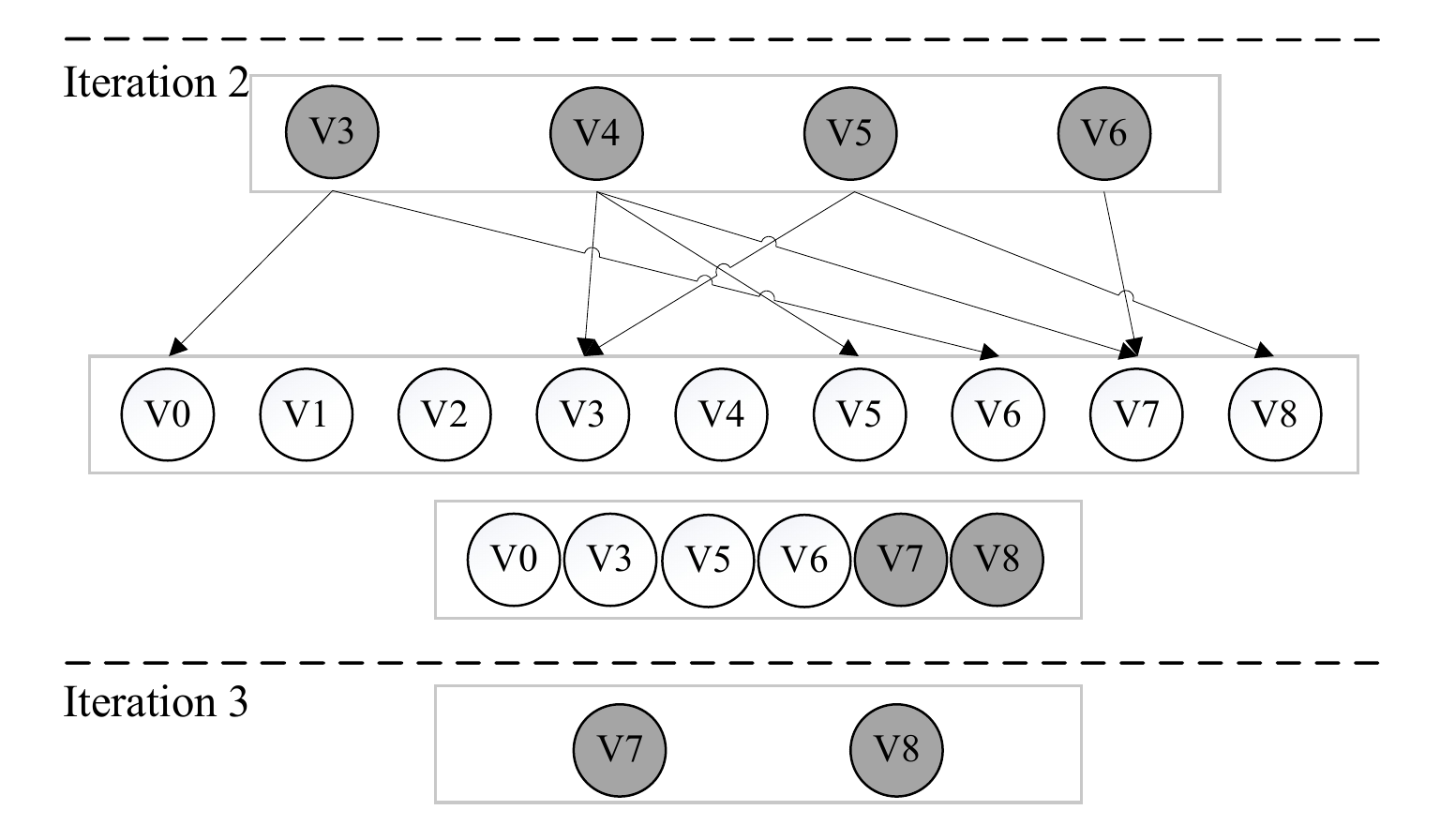}}
\caption{Push-style in Vertex-centric}
\label{fig}
\end{figure}

\subsection{\textbf{Workload Balancing}}
The degree of the vertex set follows the power law distribution so that the degree of different vertex is very different from each other. Furthermore, the different degree of vertex results in different processing parallelism. Therefore, using the same framework to process different degree of objects isn¡¯t an appropriate strategy in graph processing. To settle the processing discrepancy of different degree, we design the workload balancing strategy to resolve the problem. On the other hand, FPGAs on-chip resources is limited and every instruction generates custom circuit, so we need to consider the resources consumption while programming functions.

In the low parallelism module with vertex-centric, the number of vertex to traverse is very small and the degree is possible low because the majority of vertex only have low degree. Chewing on the limited on-chip resource, although the degree of vertexes in this module is different, we only design a uniform framework to process all vertexes of the stage. To achieve parallel processing, we design a thread group composed of 16 threads and a group takes charge of an active vertex. If the degree of vertex is less than 16, a thread group will complete processing of it. If the degree is more than 16, a group will complete processing with a few loops executing.

\begin{figure}[tp]
\centerline{\includegraphics[width=0.48\textwidth]{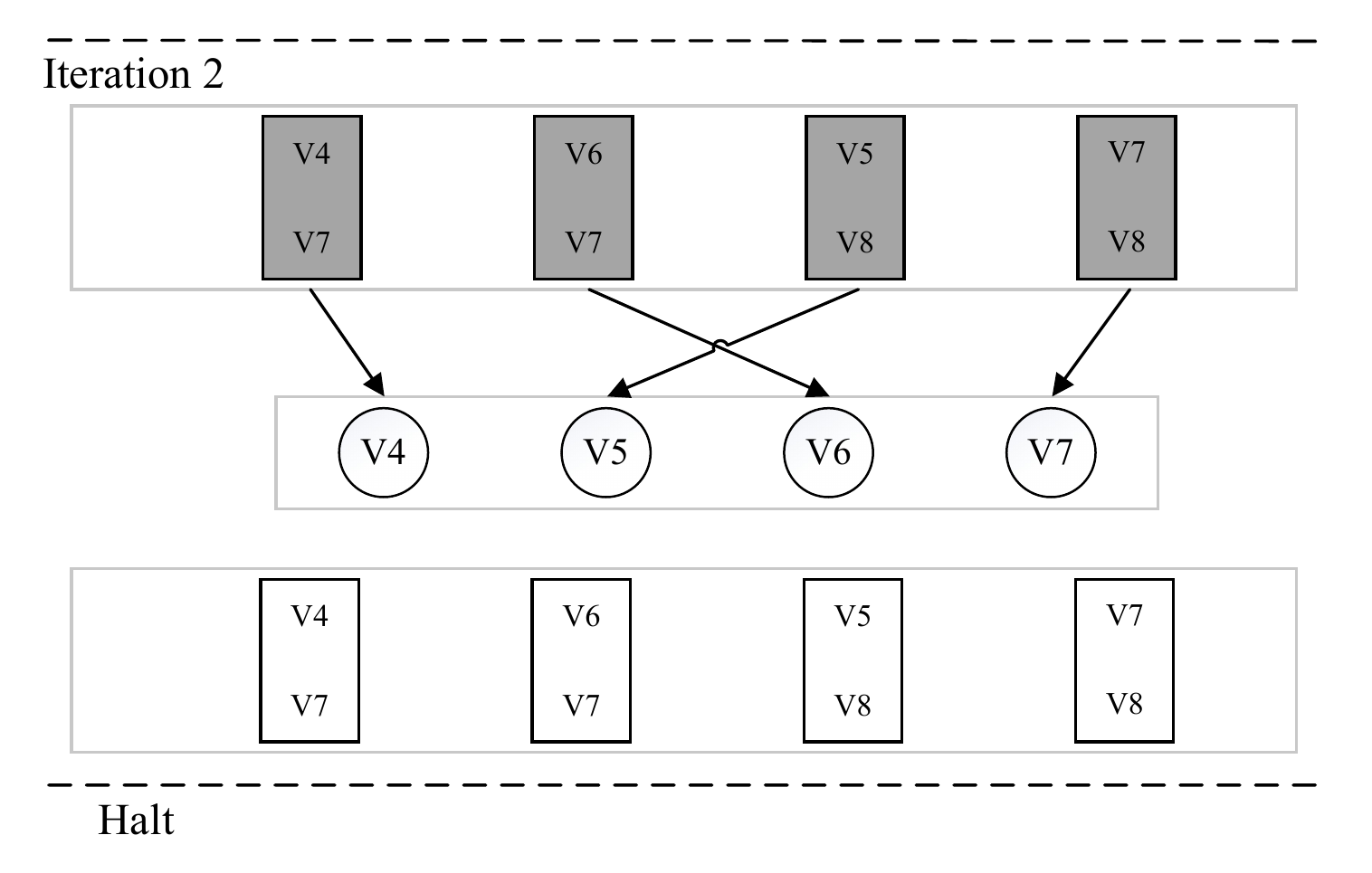}}
\caption{Pull-style in Edge-centric}
\label{fig}
\end{figure}

\begin{figure}[tp]
\centerline{\includegraphics[width=0.48\textwidth]{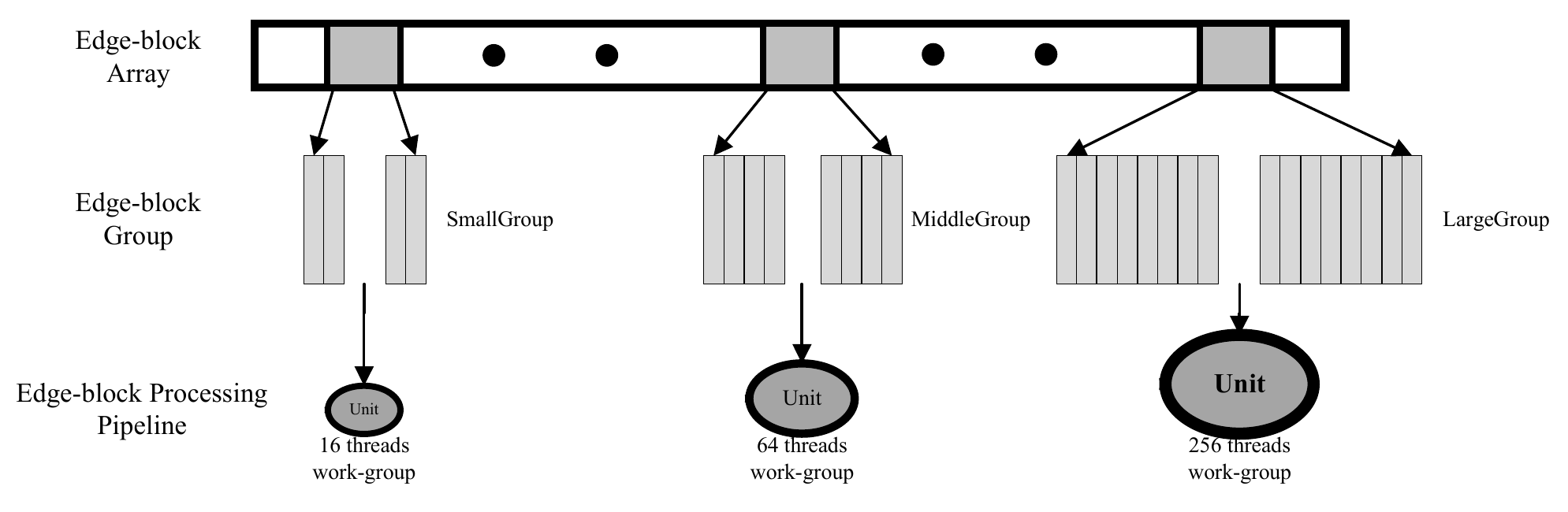}}
\caption{The Edge-block Workload Balance}
\label{fig}
\end{figure}

In the high parallelism module with edge-centric, the number of vertex to traverse is very large and almost total edge set needs to been processed. The edge set is partitioned to edge-block with different number of edges and the object to process is edge-block, rather than edges. Because of the diverse size of edge-block, using a uniform framework to process all edge-block will lead to serious imbalanced workload, resulting in extremely low performance. To process different size of edge-block, we classify the edge-block into three group, SmallGroup, MiddleGroup and LargeGroup. Particularly, the edge-block in SmallGroup has fewer than 64 edges, MiddleGroup between 64 and 2048 and LargeGroup more than 2048, shown in Figure 9. Before processing edge-blocks, we analyze the edge number of edge-block, which will be processed in the next iterations. An edge-block with certain number of edges is pushed into one of three edge-block array based on the edge number. We take full advantage of OpenCL execution model that the parallelism processing space is an NDRange with one, two or three dimensional index space composed of numerous work-items and work-items in different dimension form a work-group with the same dimension as the NDRange. We set a work-group with 256 work-items and 256 work-groups form the NDRange. At the next iteration, three edge-block groups will be processed in three kernels with different number of threads ( 1, 64 and 256) to balance the workload among different block size.

\subsection{\textbf{Process Valid Data}}
As long as processed only once, the graph data won't become active again in some graph algorithm, such as BFS. Therefore, these graph data become invalid after processed in one iteration. Similarly, some graph data are valid to be processed in the next iteration while connecting with the active data in this iteration, such as connected component (CC). In this work, we will use one bit to record the state of one vertex or edge-block to indicate the valid data which will be processed in the next iteration, forming the bitmap to indicate the entire graph data further\cite{b12}.

Based on the bitmap, we only access the valid graph data from the global memory and ignore some data if they don't take part in the processing in the next iteration. In this way, we can decrease the data transfer from the off-chip memory and improve the processing efficiency of the entire graph.

\section{\textbf{DISPATCHER MECHANISM}}
Adopting the double modules framework to graph processing, the processing state is alterative persistently and we demand to catch the dynamic executing characteristic in run-time to match the corresponding module. Consequently, the module dispatcher gains the run-time information and switch the processing module. This Section shows the detail of the dispatcher.

\subsection{\textbf{Dispatcher to the High Parallelism module}}
Some classes of graph application require the processing procedure to switch low parallelism to high, such as graph traversal algorithm\cite{b13}. The low parallelism module is program in vertex-centric and the processing mode is push-style. There are many original works offering various transforming mechanism to switch push-style to pull-style in vertex-centric. Scott Beamer et al. proposed a heuristic formula to associate the number of active vertex with the number of unexplored vertex, influenced by a tuning parameter. Ligra set the threshold to $\vert E \vert$/20 to decide when to use sparse fashion versus dense fashion. The high parallelism module program is in edge-centric and almost all edge take part in processing in this module. Therefore, the excellent position to switch to the high parallelism module is regularly when the number of vertex and edge to update is at its most.

To switch to the high parallelism module in a proper point, we adopt a dispatcher policy based on: Na indicates the number of active vertex and Ni indicates the number of inactive vertex. Na is efficient to compute when analyzing the processing state before the next iteration. When active vertexes occupy a certain percentage of the vertex set, the ¡°hub¡± vertex with very high degree is very possible to be an active vertex in the next processing iteration. Therefore, we set a tuning parameter $\alpha$ to control the active vertex threshold. The dispatcher policy to compute the point to switch follows:
\begin{equation}
\frac{N_{a}}{N_{i}} < \alpha \label{eq}
\end{equation}

Because of the "hub" vertex with a very high degree, a proper point to switch to the high parallelism module emerges when a ¡°hub¡± vertex becomes active to lead to that a large number of vertex become active in the next iteration. Consequently, while a ¡°hub¡± vertex become active, the dispatcher begins to execute the high parallelism module immediately. At the other hand, applying the switching strategy to identify the point doesn¡¯t bring on that a great many vertex and edge need to participate in processing. Therefore, when the dispatcher indicates module conversion, the system still complete this iteration in the low parallelism module. Relying on this dispatcher mechanism, we can fully utilize the performance cost to analyze the processing state and abstract the data for low parallelism stage. In addition, the system ensures that execute the processing in the high parallelism module when the processing data is the entire graph at the beginning stage.

\subsection{\textbf{Dispatcher to the Low Parallelism module}}
To undertake the role to accomplish the high parallelism processing, this module is the pivotal part to optimize the performance, which process the majority of data of graph. If a portion vertex or edge don¡¯t participate in processing, it is necessary to switch the processing module from the high parallelism to the low because the portion data will take part in the processing in this module though not bringing about the processing result, which severely damaging the processing performance. The module is in edge-centric and execute in edge-block data structure. First, we research the running processing state in the entire procedure.

\begin{figure}[tp]
\centerline{\includegraphics[width=0.45\textwidth]{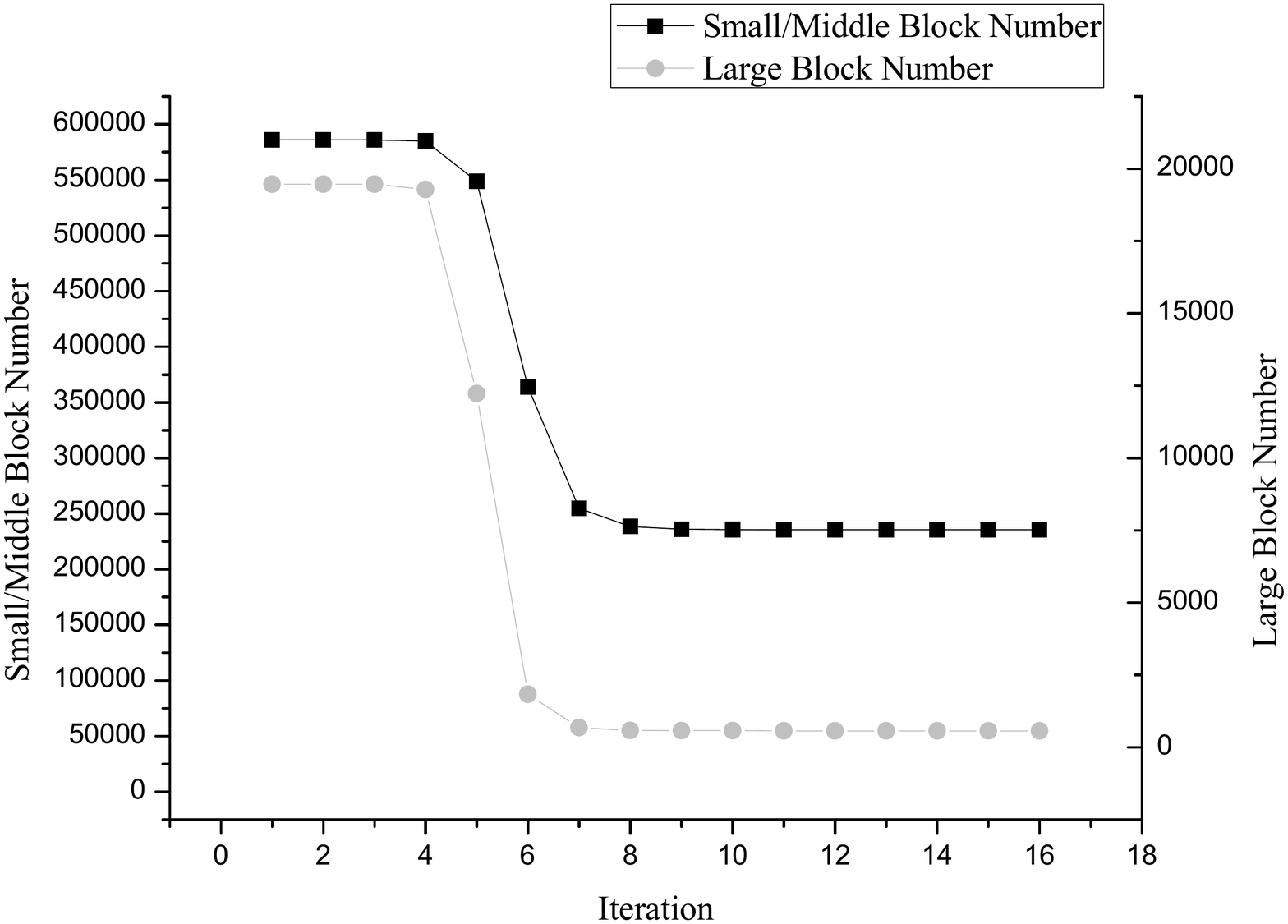}}
\caption{The Processing Characteristic of Edge-block}
\label{fig}
\end{figure}

As the executing data shows in Figure 10, the great mass of small-blocks become inactive totally and middle-blocks are the same as small-blocks, but the fraction of large-blocks become inactive and the majority of them are continuous active in the whole procedure. We can find the cause from the data structure of the large-block. As discussing in Section 2.1, the degree of vertex follows the power law distribution so the in-degree of the same destination is much possible to be low. The very high degree of destination leads to that many edges form the large edge-block. Moreover, the high degree vertex is much possible to be active continuously or become active again in some applications such as PAGERANK, and a certain part of vertexes are traversed in the final iterations or are not traversed in the entire procedure such as BFS. Therefore, the large-block is much possible to maintain active state in the whole processing. In the whole processing procedure, the edge of the same destination in block is much potential to become active or active again because of active neighbors, especially middle-blocks and large-blocks. When edges of the great mass of small-blocks become inactive, a quantity of middle-block and the majority of large-block are active. However, the number of small block is much more than the number of small and large.

Ground on the running characteristic of edge-block\cite{b14}, we adopt the dispatcher policy based on: $F_{l}$ indicates the access flag number of the large block, $N_{l}$ indicates the number of the large block, $N_{b}$ indicates the number of the sum of the small and the middle, and $N_{a}$ indicates the number of active small and middle block. In some graph applications such as PAGERANK, the entire edge set is active in the initial iteration and edge blocks become inactive along with the processing procedure. In others such as BFS, edge blocks become traversed gradually level by level, but some edges are not traversed in the whole procedure. In addition, we set a parameter $\beta$ to control the active edge-block threshold and another parameter $\gamma$ to record the access state of the large edge-block. Consequently, we compute the point to switch to the low parallelism module following:

\begin{equation}
\frac{N_{a}}{N_{b}} > \beta \label{eq}
\end{equation}

\begin{equation}
\frac{F_{l}}{N_{l}} > \gamma \label{eq}
\end{equation}

The switching policy has two conditions. When the formula 2 has been established, once the formula 3 is established, the dispatcher switches the processing module. When the formula 2 is established but the formula 2 hasn¡¯t been established, the processing still executes in the original module and will switch to the low module in the next iteration. Depending on the policy, the processing module can switch to the low in time without unnecessary operators to analyze the whole state of processing and utilize the state analysis overload while the processing state possibly suits the low module but the performance is still almost the same in the high module.

\section{\textbf{EDGE-BLOCK MODULE}}
The edge-block is the key mechanism to optimize the high parallelism state, which processes the majority of the graph data. The edge-block can realize the stream parallelism easily and avoid extra overload to process inactive edges. This section shows the detail of the edge-block.

\subsection{\textbf{Data Structure}}
The input of this module is an unordered edge list without a great deal of pre-processing overload to generate complicated date structure\cite{b15}. The edge-block is the edge group of the same destination vertex. The least size of type of data in OpenCL is the char type and the pipeline depth of the kernel logic is always the power of 2. Therefore, the number of destination vertex in one block is a certain number of the power of 8 so that we can use a certain number of bytes to record one whole block such as a char type. The vertex member to one edge-block is the sequential position in the unordered edge list so that we don¡¯t need to sort the edge list and the necessary overload to generate edge-block is extremely low. When generating the edge-block set, we will gain the number of the edge of the block so that we can distinguish the size of the edge-block at the same time. Based on the size of edge-block, we generate an index array to record the number of edge of different block. In order to identify the position in edge list, we also use an index array to record the starting position of edge-block in the edge list. The Figure 11 illustrates the data structure of the edge-block of the graph of the Figure 1.

The destination vertex set of an edge-block is a subset of the vertex set of graph so that edge sets of different edge-block are disjoint and total edge-blocks equals the edge set of the entire graph. When initializing the data structure of edge-block, the destination vertex set is classified into subsets depending on the sequential position. Edges are also partitioned by different destination vertex sets\cite{b15}. The edge-block structure maintains fixed in the entire procedure but the number of block set is mutative during the processing when some edge-block become inactive. After every iteration, the dispatcher analyzes the processing state and generates the active block array to the module to prepare the next iteration. A flag array records the access state of large edge-block state. The block array and the flag array are mutative after every iteration\cite{b16}.

The partition size is also an important factor for processing performance\cite{b17}. The scale of graph is fixed. Therefore, if the number of edge-block is large, mostly edge-blocks is small ,and a small block can store in the on-chip BRAM easily to make full use of the BRAM to realize the high-speed access to boost performance. However, the large number of edge-block results in that the majority of edge block only own a very few edges because of the power law distribution so that the majority of precious BRAM maintains idle state during processing. If the number of edge-block is small, the dispatcher will analyze the processing state and dispatch workload easily. Whereas, if the number is too small, every function unit will take much time to process one block in several iterations, leading to reduce performance greatly.

We try to find the upper bound of the size of the power of the edge-block\cite{b18}. We indicate the entire graph data with G, the pipeline depth with D and the size of power with n. In order to saturate the pipeline, the size of a edge-block need to be larger than the pipeline depth. As the majority of the degree is very low and almost equals to 1, we consider the degree of one vertex as 1 for simplicity. Consequently, we can get such a formula: \begin{math} D < \frac{G}{8^{n}\times P} \end{math}. Furthermore, we can get the formula:
\begin{equation}
n < \sqrt[8]{\frac{G}{D \times P}}\label{eq}
\end{equation}

\begin{figure}[tp]
\centerline{\includegraphics[width=0.42\textwidth]{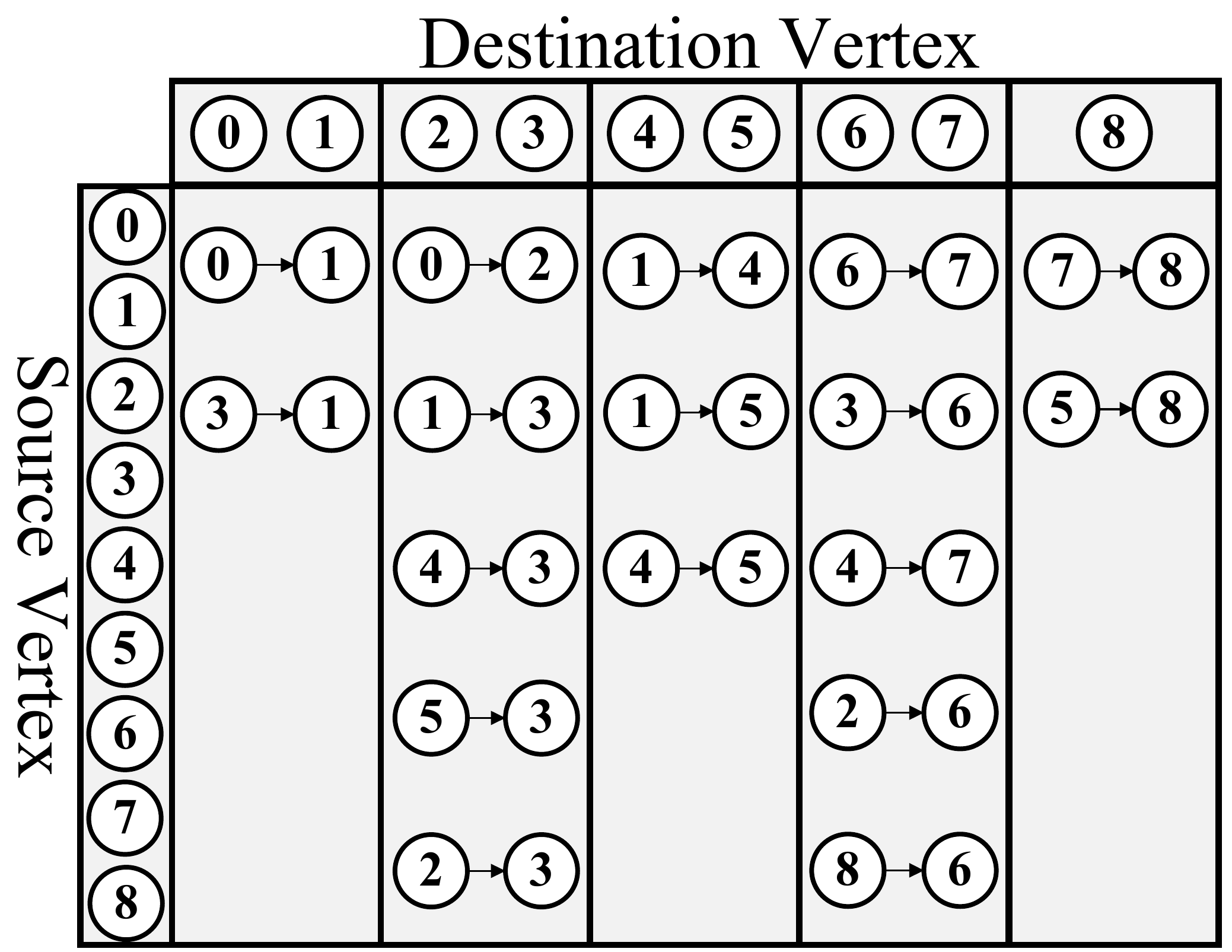}}
\caption{Edge-block Data Structure}
\label{fig}
\end{figure}

\subsection{\textbf{Pipeline Parallel}}
To realize workload balance, we classify the edge-block into three class depending on the block size and process different size of block by different thread group, showed in Section 3.4. FPGAs utilize on-chip resource to generate custom circuit to pipeline processing in different function units. In addition, Altera OpenCL framework is pipeline parallelism. The key factor to realize high parallel processing is to access edges and set instruction circuit reasonably.

Based on the power law distribution, the number of edge-blocks with less than 64 edges, called small block, is very large. Therefore, the time to process small edge-block need to be as small as possible. In the other hand, the pipeline depth is certain size in one pipeline circuit. We utilize 1 thread to process a small block and the thread processes sequent edge in pipeline mode. Because of the block size, the thread group uses less than 8 loop to finish processing one small block. Making full use of pipeline processing and several pipeline circuits, thousands of edges can be processed at the same time. A short loop step finishes processing a block in very little time and huge parallelism can parallel process a large number of edges at the same time. Moreover, the order to process edges is sequential without random access from DRAM.

The number of middle blocks, whose number of edges is between 64 and 2048, is still large but the number of edges is much larger than the small block. To control the loop times for processing, we utilize 64 threads to process one block. Because of the size of the middle block, the group can finish processing a block in less than 32 loop. During the size of the middle block becomes larger and larger, the number of it becomes smaller and smaller. Therefore, most of the middles blocks can been processed in a few loops. Furthermore, the executing state of different work-group is independent in Altera OpenCL framework so that the larger number of loops doesn¡¯t influence the small group. Therefore, the middle group and the small group process respective edge-blocks simultaneously and the less number of block helps the middle group to catch the entire processing time so more number of loops doesn¡¯t decrease the entire performance.

The large edge-block, whose number of edges is more than 2048, is the key role to realize high processing performance because almost edge-blocks are connected to them. The block is very different form the small and middle because the upper threshold of the edge number doesn¡¯t exist and the number of the block is much less than others. To process a large block, we utilize a work-group composed of 256 threads to parallel compute the state of edges. In every loop, 256 threads process respective edge simultaneously and a work group need to take more than 8 loops to finish processing one block. Therefore, the processing pipeline is full in the whole procedure. We use a few memory in every work-item as an edge cache to store edges processed in the next loop, which short the Initiation Interval of different loops to speed up the pipeline performance. Because of the finite BRAM, if one work-item need to consume one BRAM, one work-group need to consume hundreds of BRAM. Furthermore, an NDRange to process the large block will consume all BRAM, but this processing mode can¡¯t implement in FPGAs. Consequently, we limit the cache size to be 64 bytes so that the cache is composed of registers in work-item. In addition, the access speed of register is quicker than BRAM so the pipeline processing speed can gain the best performance.

The three classes of edge-block are respectively processed in different size of function units and three work-group pipeline can execute simultaneously through three command queues in Altera OpenCL framework. The parallel framework of FPGAs is pipeline mode and the edges stored in DRAM is sequential so that the data reading and writing mode is sequential access mode. As showed in Section 4.2, the edge-block owns one uniform state to indicate dispatcher to extract the active blocks for the next iteration. If there is not any edge in block is active, the state of the block will become inactive and don¡¯t participate in processing in the next iterations. Consequently, the edge-block dispatcher mode avoids the random access and realizes the high parallel processing.

\subsection{\textbf{Pipe Transfer}}
One edge-block is an independent cell in the edge-centric processing stage, preventing the system from processing the entire graph in each iteration. The system transfers edge-block from one unit to another to complete respective processing task. The bandwidth between the chip and the off-chip memory is finite and on-chip processing units always doesn't realize 100$\%$ utilization of it to access data from the global memory. Furthermore, to achieve data communication between different units through the global memory is a very high overload task and will consume a great deal of time to guarantee synchronization and correctness. The pipe is the one first input first output (FIFO) buffer, allowing kernels to communicate with each others directly through the BRAM. Without access with DRAM, it is very efficient to pass data between kernels and synchronize processing procedure of different units with low latency.

\begin{figure}[tp]
\centerline{\includegraphics[width=0.40\textwidth]{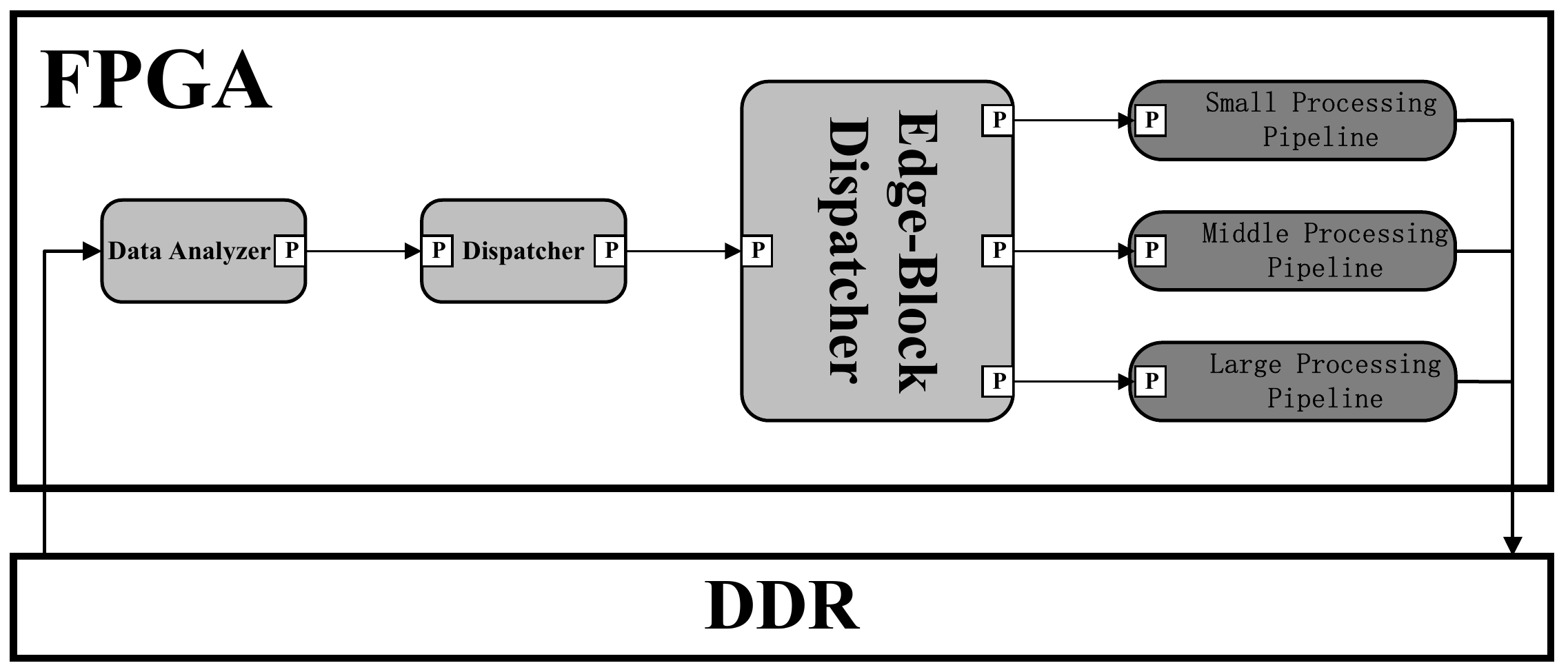}}
\caption{Edge-block Transfer}
\label{fig}
\end{figure}

The edge-block dispatcher gathers edge-blocks to analyse the edge array size and dispatchs blocks to the processing pipelines according to the size. Thanks to the independency of the block, the system can dispatch blocks into the corresponding pipeline to achieve custom processing. Because of one input site and one output site, a pipe is custom to transfer a class of edge-block. Therefore, the dispatcher transfers blocks into corresponding processing pipeline using pipe buffer through BRAM, shown in Figure 12. The processing pipelines gain blocks from respective pipes to complete processing sequentially and terminate after receiving the signal to halt.

In previous work, the dispatcher analyses the graph data and prepares a serious of data for processing and the processing unit acquires data from the global memory to complete graph processing. It is seldom to parallel the double procedures because there are a large number of load and store operations. Using pipe to transfer data between different kernels prevents kernel from communicating with each others through the global memory and these kernels can parallel execute to complete graph processing without stall. We can overlap the dispatcher and the processing procedure to save the time of dispatching and data transfer between BRAM and DRAM. In addition, the FIFO transfer rule can guarantee the data consistency between units without more overhead or synchronization operations.

\section{\textbf{EVALUTION}}
Based on the dispatcher and system design mechanism, we conduct a serious of experiments to evaluate performance and measure optimization effect. In addition, we compare the performance distinction processing on FPGAs with processing on CPU. We also compare the performance of our system with the state-of-the-art systems in this section.

The system is implemented in more than 2500 lines of OpenCL host code and 800 lines of kernel code. The OpenCL host code is compiled with G++ 5.4.0 on Ubuntu 16.04 with the optimization flag of O2 and the kernel code is compiled with aocl 16.0.0.211 based on OpenCL 1.0. All of the experiments related to FPGAs in this paper were established on a Linux server with 2$\times$2.4 GHz Intel 14-core, hyperthreaded E5-2680 v4 Xeon processors, 1 TB of main memory and an Intel Arria 10 GX 1150 FPGA with up to 6.62 MB on-chip BRAMs, 16GB on-board memory. The FPGA chip provides 427,200 K Adaptive logic modules (ALMs) and 1,708,800 registers. The peak bandwidth of off-chip memory is 30GB/s. We can verify the processing characteristic of the system through Altera aocl report GUI tools. Because of the OpenCL SDK conflict, we ran the experiments executed on CPU in a single machine with 3.30 GHz Intel(R) Core(TM) i5-4590 4-core CPU, 256 KB L1 cache, 1.0 MB L2 cache, 6.0 MB L3 cache and 8GB RAM. The OpenCL code is compiled with Visual Studio 2017 on Window 10.0.

\subsection{\textbf{Graph Algorithms and Data Set}}
We evaluate the performance of our system using the following three algorithms:
\begin{itemize}
\item \textbf{Breadth-First Search (BFS).} Traverse a graph from a source vertex to all other vertex along with the degree of traversed vertexes in each iteration.

\item \textbf{PAGERANK (PR).} Compute the PageRank value for all vertex according to the linking between vertexes of a specified graph using a rank computing algorithm.

\item \textbf{Weakly Connected Components (WCC).} Consider a specified graph as a undirected graph and compute the number of connected components in the graph by some path, ignoring direction.
\end{itemize}

The graph data sets and their acronyms using in our experiments to evaluate the processing performance are shown in table \uppercase\expandafter{\romannumeral1}. The four graphs is real graphs selected from . The four graphs are all online social network with totally different scale and degree distribution. These graphs have be preprocessed in the host to generate the CSR array and the edge-block array. These graph data is transferred to the FPGA board and stored in the DRAM. The CSR array and the edge-block array are generated by the degree of the source and destination so that the preprocessing overhead can be restricted in \emph{O}($\vert$\emph{E}$\vert$) time to traverse the entire edge set. The graph processing system can process the unordered edge list without more preprocessing apart from overhead above-mentioned.

\begin{table}[!t]
\renewcommand{\arraystretch}{1.6}
\caption{Graph Datesets and Acronyms}
\label{table_example}
\centering
\setlength{\tabcolsep}{5pt}
\renewcommand{\arraystretch}{1.5}
\begin{tabular}{c|cccc}
\hline
Dataset& Vertexes& Edge& Diameter& Type\\
\hline
soc-Epinions (EN)& 0.08M& 0.51M& 14& Directed\\
com-Youtube (YT)& 1.16M& 2.99M& 20& Undirected\\
soc-Pokec (PK)& 1.63M& 30.6M& 11& Directed\\
Live-journal (LJ)& 4.85M& 69.0M& 16& Directed\\
\hline
\end{tabular}
\end{table}

\begin{table}[t]
\renewcommand{\arraystretch}{2.5}
\caption{Resource Utilization}
\label{table_example}
\centering
\setlength{\tabcolsep}{4pt}
\renewcommand{\arraystretch}{1.5}
\begin{tabular}{c|cccccc}
\hline
Algorithm& LU& ALUT& Registers& BRAM& DSP& Clock rate \\
\hline
BFS& 87\%& 36\%& 52\%& 95\%& 9\%& 215 MHz\\
WCC& 95\%& 37\%& 58\%& 97\%& 10\%& 189 MHz\\
PR& 98\%& 39\%& 59\%& 97\%& 13\%& 178 MHz\\
\hline
\end{tabular}
\end{table}

\subsection{\textbf{Execution Performance}}
\begin{table}[b]
\newcommand{\tabincell}[2]{\begin{tabular}{@{}#1@{}}#2\end{tabular}}
\renewcommand{\arraystretch}{2.5}
\caption{Graph Datesets and Acronyms}
\label{table_example}
\centering
\setlength{\tabcolsep}{5pt}
\renewcommand{\arraystretch}{1.5}
\begin{tabular}{c|c|cc}
\hline
Algorithm& Graph& \tabincell{c}{Execution \\ time(s)}& \tabincell{c}{Throughput \\ (MTEPS)}\\
\hline
\multirow{4}{*}{BFS}& EN& 0.006& 85\\
&YT& 0.028& 107\\
&PK& 0.152& 201\\
&LJ& 0.395& 175\\
\hline
\multirow{4}{*}{WCC}& EN& 0.010& 102\\
&YT& 0.037& 162\\
&PK& 0.164& 373\\
&LJ& 0.370& 370\\
\hline
\multirow{4}{*}{PR}& EN& 0.003& 170\\
&YT& 0.043& 70\\
&PK& 0.244& 125\\
&LJ& 0.619& 111\\
\hline
\end{tabular}
\end{table}

The detail of resource utilization is shown in table \uppercase\expandafter{\romannumeral2}. To adapt to different scale of graph, we use 4 bytes integer to represent the index of vertex and edge. The space of DRAM is 16GB and 4 bytes integer can represent the max number is 4G so that we can process the majority of nature graphs. In BFS, we use 1 byte char type to represent the depth of a vertex and the bitmap to represent the access state of the vertex set. In WCC, we use 4 byte integer to represent the mark and the bitmap to represent the change state of the vertex set. In PR, we use 4 byte float type to represent the rank value of vertex and the bitmap to represent the convergence state of vertex set.

To evaluate the processing performance of the double module graph processing, we ran three graph algorithm ( BFS, PR and WCC) on four graph ( EN, YT, PK and LJ). The executing result is shown in table \uppercase\expandafter{\romannumeral3}.

\subsection{\textbf{Optimization Efficiency}}
\subsubsection{Double Module Processing}
The degree of vertexes in nature graph has power law distribution, resulting in two parallel processing procedure. In this paper, we propose two processing module to match the running characteristic of double state. We implemented the BFS on four graphs using different processing mechanism include the vertex-centric processing (VC), the hybrid of push-style and pull-style processing (VCH), the edge-centric processing (EC), the combination of vertex-centric and edge-centric processing(ECH), the edge-block processing (EB) and the double module processing (DM).

\begin{figure}[b]
\centerline{\includegraphics[width=0.42\textwidth]{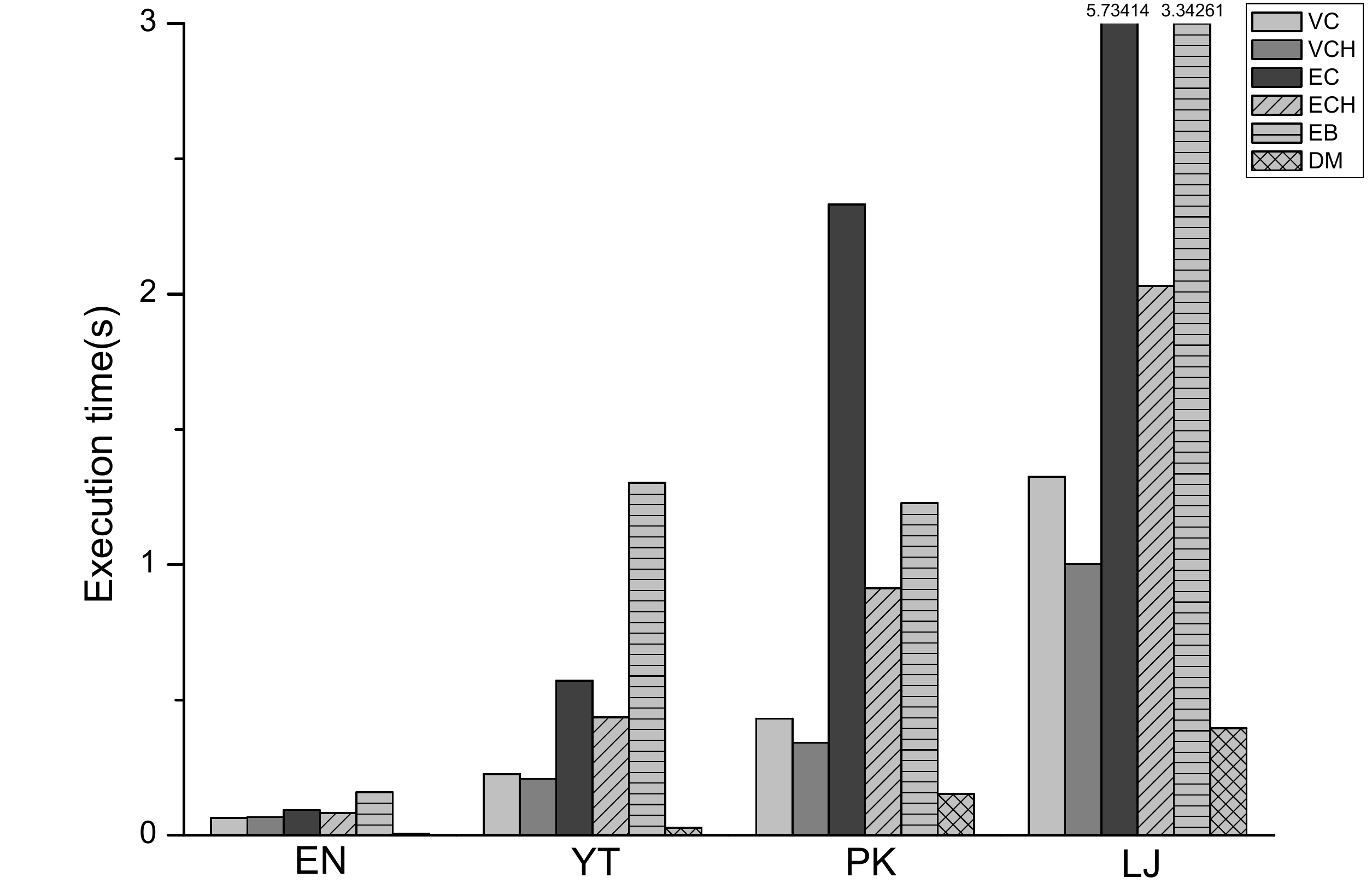}}
\caption{Processing Performance of Different Architecture}
\label{fig}
\end{figure}

Figure 12 shows that the double processing mode mechanism can improve processing performance such as VC and VCH. Except for processing on EN, the time consume in VCH is 1.2x performance improvement comparing to VC because EN is a small scale graph so that we don't spend too much time to complete the processing procedure. Therefore, the time to switch the processing mode occupies a certain part of the total time even if the time to change mode is very little. The total of time to processing four graph in VC is much more than VC and VCH because VC consumes the same time in every iteration although the graph data decreases after each iteration. To overcome these cause, ECH adopts the vertex-centric mode to processing some iterations which graph data to processing is small and gain 2$\times$ performance promotion. The overload to complete the processing procedure in ECH is still more than VCH because the graph data decreases fast while processing in the high parallel procedure. In this paper, we adopt the edge-block to complete the graph processing and the performance is worse than EC. The edge set streams into CUs sequentially in VC and CUs require spending a certain of time to compute the edge array position, resulting in many transfer stall. However, the time to finish in EB is still much less than EC. As shown in Figure 12, the time to complete the entire procedure in MD is almost 3$\times$$\sim$25$\times$ times less than others, which adopts in workload and pipe transfer in edge-block processing stage. The experiment result proves that the double module processing mechanism can immensely raise the performance processed on FPGAs.

\subsubsection{Workload Balance}
The degree of vertex in nature graphs obeys power law distribution, leading to that the majority of edge-block have a short edge array, the minority have long edge array and a few huge edge-block own very long one. To process different scale of edge-blocks, we design different size of work-groups to processing corresponding edge-blocks to prevent a work-item from spending too much to process one edge-block. We execute the BFS on four graph in the single edge-block processing mode using three different workload allocation mechanism, the single processing work-group type, double type and three group type.

\begin{figure}[b]
\centerline{\includegraphics[width=0.36\textwidth]{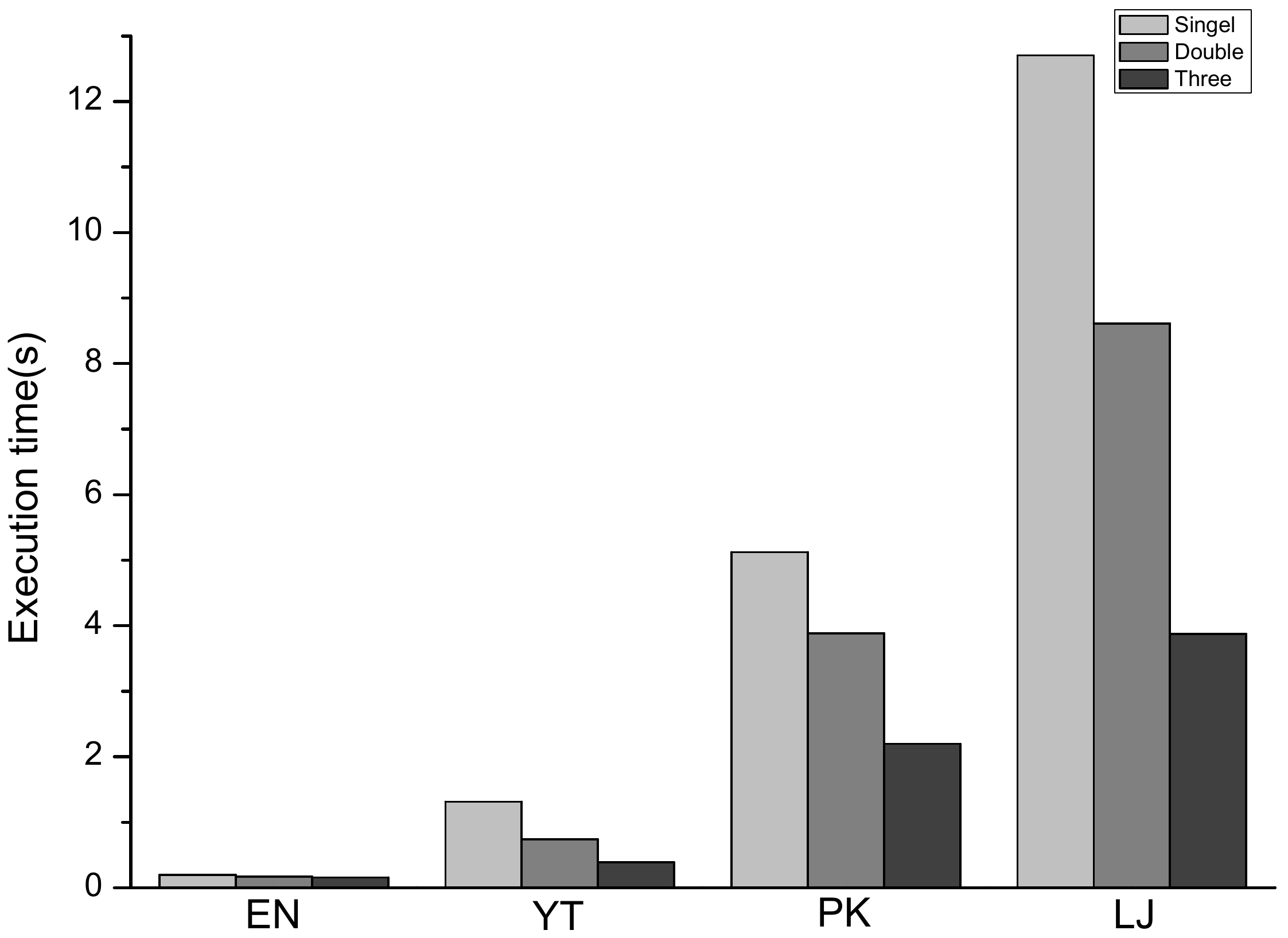}}
\caption{Workload Balance}
\label{fig}
\end{figure}

Along with better workload balance, the time to complete processing the entire procedure become less and less, as shown in Figure 13. A few hub vertexes have very huge degree while the majority of vertex with low degree. The total processing time is dependent on the one unit that consumes the most time to finish. If the same size of work-group take charge of double scale of graph data, the one group squares to spend hundreds of times to finish than the other one while encountering the huge block. Compare to the single processing work-group type, the double type can realize 1.5$\times$ performance promotion and the third type can realize 1.2$\times$$\sim$4$\times$ performance promotion. The three group type gains better performance because it can implement more fine grain balance.

\subsubsection{Pipe Transfer}
The bandwidth between the on-chip memory and the global memory decreases the data transfer efficiency from one kernel to the other. On the other hand, a kernel is difficult to make full use of the 100$\%$ global memory bandwidth available. Particularly in graph processing, there is a large quantity of data access to update the graph state. We adopt the edge-block dispatcher mechanism to realize the edge-centric processing procedure in order to utilize the pipe to transfer graph data between kernels, getting ride of the lengthy procedure between the on-chip BRAM and the off-chip DRAM. In addition to some symbols introduced in Section C.1, the CVCH indicates adopting pipe into the VCH and the CDM is the same meaning.

\begin{figure}[t]
\centerline{\includegraphics[width=0.35\textwidth]{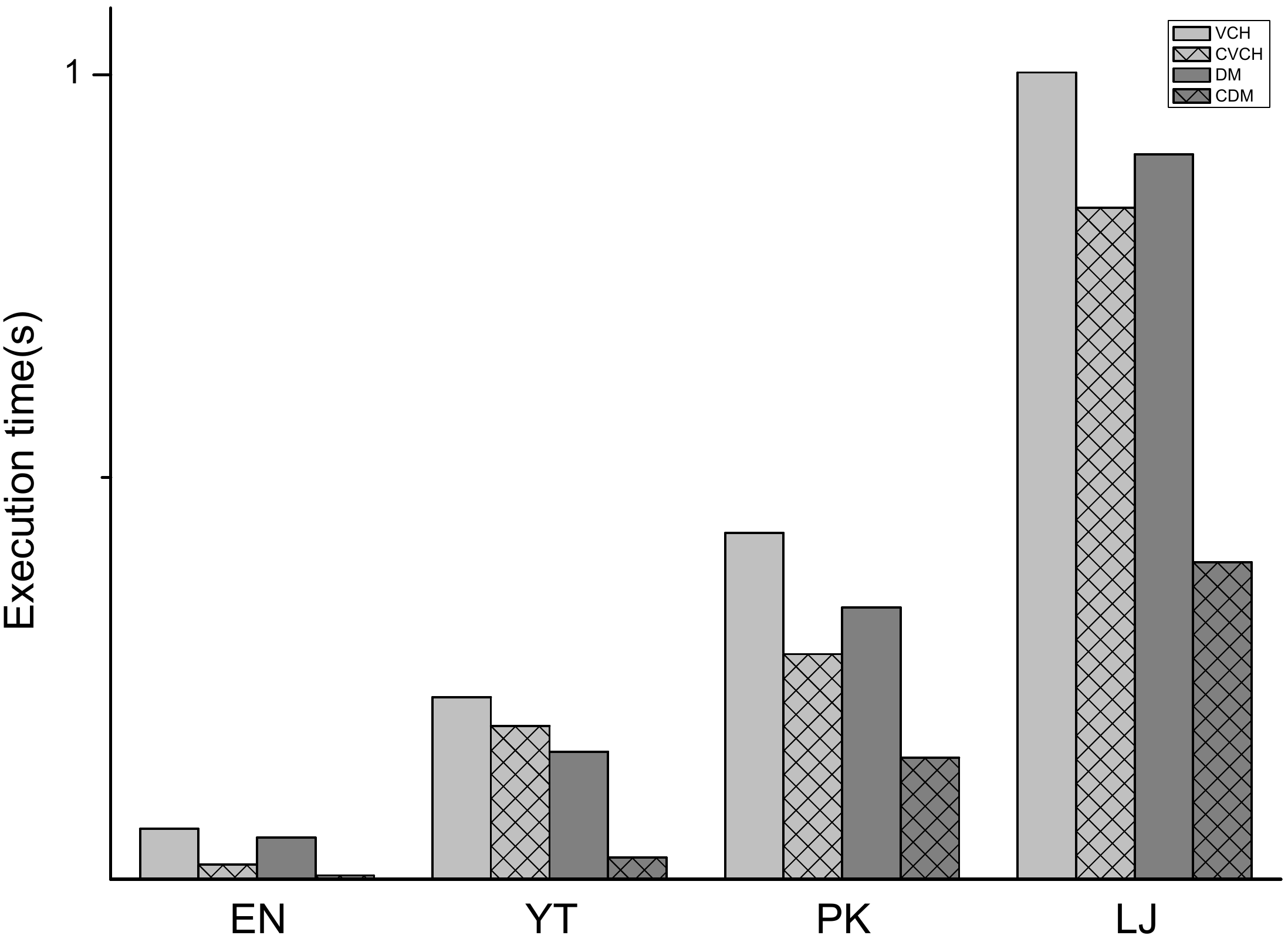}}
\caption{Pipe Transfer}
\label{fig}
\end{figure}

Applying the pipe to the processing mechanism, we can gain 1.15$\times$$\sim$3$\times$ performance promotion in VCH and 2$\times$$\sim$8.6$\times$ performance promotion in DM, shown in Figure 14. The cause of the performance promotion efficiency in CDM better than in CVCH is that the CDM designs more kernel to realize workload balance which generating more data transfer between kernels. What's more, Figure 14 indicates that the edge-block dispatcher achieves better performance than the VCH or the CVCH while processing YT.

\subsubsection{The Size of Edge-block}
The edge-block is the dispatcher cell in edge-centric processing state. The size of edge-block is related to the block switching frequency and processing performance. We implemented the BFS on four graph and change the size of edge-block, a certain number of the power of 8, to show the processing diversity. To let the experiment result visualized better, we process the EN and the YT only in the edge-block processing to superimpose over all iterations because of the small size of double graph, while the PK and the LJ implement in CDM.

\begin{figure}[t]
\centerline{\includegraphics[width=0.36\textwidth]{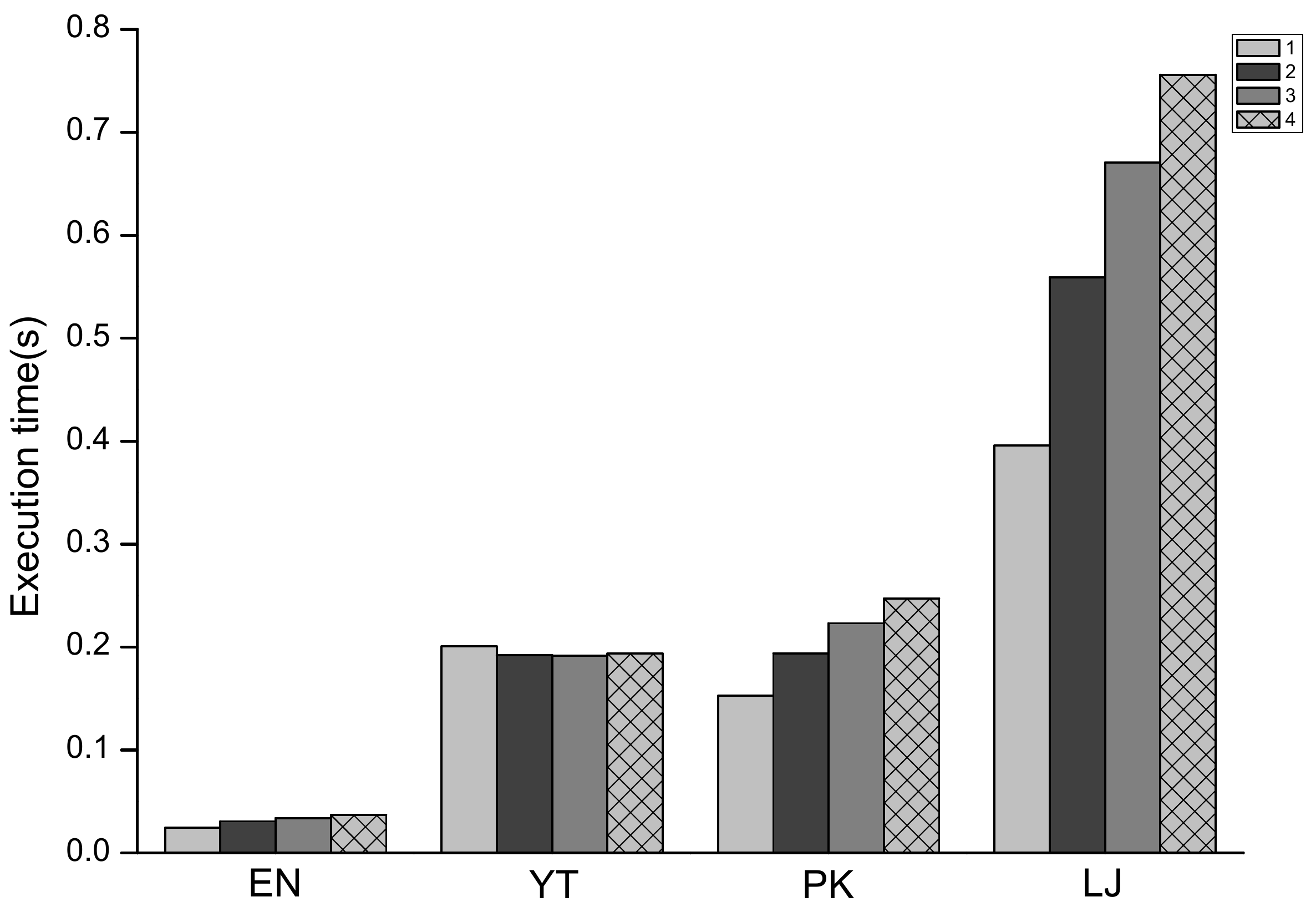}}
\caption{Block-size Influence}
\label{fig}
\end{figure}

Figure 14 shows that the block size can influence the processing performance obviously. While processing in EN, YT and LJ, smaller size of blocks can gain 1.25$\times$$\sim$1.9$\times$ performance compared to others because of vertexes with low degree in these graphs occupy a large part. However, the block consisted of 32 vertexes gains the best performance while implementing in YT. The cause of this result is that the proportion of vertex with low degree is smaller than the other three graphs.

\subsection{\textbf{Comparison to CPU processor}}
We execute BFS in the optimized VCH mode on CPU to gain the performance. CPUs are a kind of processor composed of a certain number of processing units and these units process the same computing task meanwhile with different processes to realize parallelism. Consequently, the vertex-centric processing mode is suitable for it and the hybrid mode can improve performance apparently in graph processing. Even so, the double modules processing system gain better performance compared to process in CPU. While processing in the same graph, our system can gain 1.41$\times$$\sim$5.0$\times$ performance promotion compared to CPU.

\subsection{\textbf{Comparison with Other Systems}}
We compare performance of this work with state-of-the-art systems on FPGA in Table \uppercase\expandafter{\romannumeral4}. Our system achieves 2.38$\times$ speedup to the GraphOps while implementing PR in PK. Compared to ForeGraph, we can catch up with the processing performance and outperforms it to implement BFS in LJ. The performance implemented in YT is worse than ForeGraph because the graph is a small graph and the transfer time from the global memory consumes more time compared to the processing time.

\begin{table}[t]
\newcommand{\tabincell}[2]{\begin{tabular}{@{}#1@{}}#2\end{tabular}}
\renewcommand{\arraystretch}{2.5}
\caption{Graph Datesets and Acronyms}
\label{table_example}
\centering
\setlength{\tabcolsep}{5pt}
\renewcommand{\arraystretch}{1.5}
\begin{tabular}{c|c|c|c|c|c}
\hline
System& Algorithm& Graph& Performance& Ours& Improvement\\
\hline
GraphOps& PR& PK&37 MTEPS& 125 MTEPS& 2.38$\times$\\
\hline
\multirow{6}{*}{ForeGraph}& \multirow{2}{*}{BFS}& YT& 0.010 s& 0.028 s& 0.36$\times$\\
 & & LJ& 0.452 s& 0.398 s& 1.14$\times$\\ \cline{2-6}
 &\multirow{2}{*}{WCC}& YT& 0.030 s&0.037 s& 0.81$\times$\\
 & & LJ& 0.307 s& 0.370 s& 0.83$\times$\\ \cline{2-6}
 & \multirow{2}{*}{PR}& YT& 0.016 s& 0.043 s& 0.37$\times$\\
 & & LJ& 0.578 s& 0.619 s& 0.93$\times$\\ \cline{2-6}
\hline
\end{tabular}
\end{table}

\begin{figure}[t]
\centerline{\includegraphics[width=0.36\textwidth]{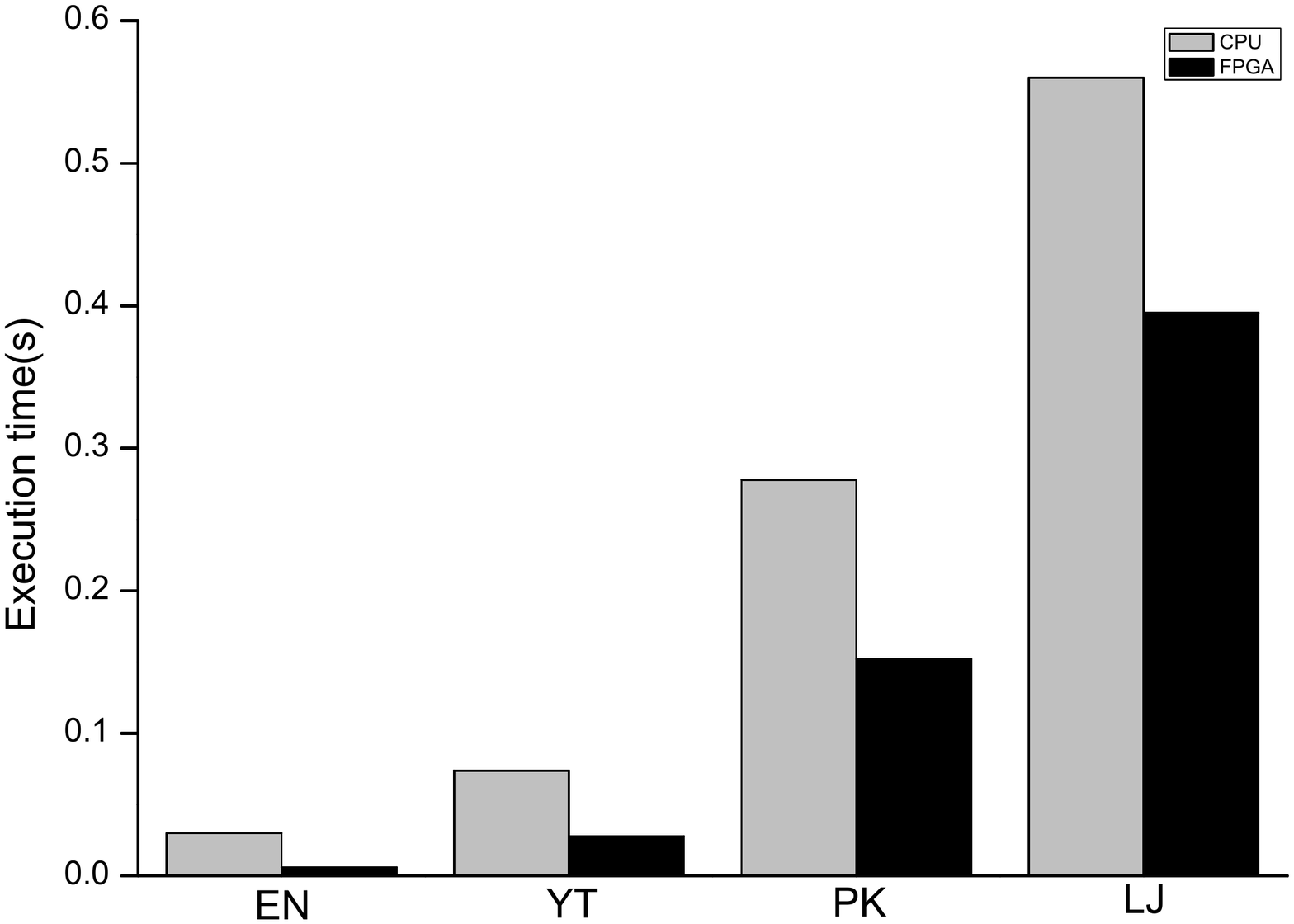}}
\caption{Performance Comparison Between CPU And FPGA}
\label{fig}
\end{figure}

\section{\textbf{RELATED WORK}}
There are several previous work designed for large-scale graph processing system based on FPGAs. GraphOps generates a modular hardware library to construct accelerators for graph analytics applications using High-level synthesis (HLS). GraphOps is built upon a dataflow platform based on FPGAs to establish a general hardware system to target a serious of analytics processing, hiding the low-level hardware dispatcher and implementation detail. To solve the random access of GAS procedure, GraphOps proposes a strategy to adopt a locality-optimized array storing the properties associated with neighbors to improve locality. Shijie et al. design a large-scale graph processing system on a single FPGA board with edge-centric computing principles based on partitioning. Shijie et al. proposes an optimized data layout to sort the edge partition based the indexes of the destination vertex or source vertex to optimize non-compulsory row-conflict. They also adopt a BRAM activation schedule to reduce memory power and generates a pipeline parallel framework to saturate the external memory bandwidth. Considering finite on-chip resource, ForeGraph constructs a large-scale graph processing system based on the multi-FPGAs architecture. ForeGraph partitions the graph edges and vertexes according to index using hash function and stores different partitions in respective FPGAs boards. Graphs are partitioned into intervals and shards and ForeGraph adopts destination-first replacement strategy to update on-chip data. Moreover, ForeGraph adopts a serious of techniques to fully utilize off-chip bandwidth and reduce communication overload.

Based on the highly pipeline parallel processing capability, a serious of work to accelerate graph algorithm emerge constantly. FPGP generates a streamlined vertex-centric graph processing framework to study the executing characteristic of breadth-first search (BFS). FPGP adopts the interval-shard structure to execute the data updating strategy. FPGP is adaptable to different graph algorithms and can scalable to multi-FPGA using a Shared-vertex Memory (SVM). Jialiang Zhang et al. present a graph processing system based on software/hardware co-design and co-optimization on a FPGA-HMC platform. Jialiang Zhang et al. modify the level-synchronize BFS to leverage the advantages of HMC of low random access latency and high parallelism. They propose a Map-Reduce-like FPGA-HMC graph processor to fully exploit the potential of FPGA and HMC. After analysing the potential bottlenecks, they propose a two-level bitmap scheme to reduce the unnecessary HMC access to optimize performance. Based on the work represented above, Jialiang Zhang et al. adopt hybrid graph traversal algorithm in FGPA-HMC graph processor. They developed two graph traversal algorithm optimization technique, degree-aware adjacency list reordering and degree-aware vertex index sorting, which the former can reduce the number of redundant edge checks and the latter can plot a strong correlation between vertex index and data access frequency. They further developed two platform-dependent degree-aware optimization techniques: degree-aware
data placement and degree-aware pointer compression to further accelerate the external memory access.

\section{\textbf{CONLUSION}}
The paper proposes a dispatcher mechanism and runtime environment for graph processing based on the double module design on a single FPGA board. The power law distribution of the vertex degree in nature graphs lead to double different characteristic processing procedures and there is no runtime connection between iterations in graph processing. Therefore, we design a double module processing mechanism to process the double different parallelism stage. FPGAs generate the pipeline to achieve parallel data processing and the data streams into the pipeline sequentially. There is a great deal of random access in graph processing and the mechanism to stream edge set into pipeline can achieve sequential access. We combine edge-centric mode with vertex-centric mode to match the processing characteristic in different stage. To parallel the entire processing procedure, we propose the edge-block mechanism to achieve data transfer between units by using pipe. The experiment result has demonstrated that the double module processing can catch up with or achieve better processing performance compared to previous FPGA-based systems. As far as I know, this paper is the first work to combine the push-style and the pull-style to achieve graph processing on FPGA and generates double processing module combining the edge-centric mode and the vertex-centric mode in an accelerator to match different processing characteristic. The finite on-chip resource extremely restricts the processing performance, especially BRAMs. Therefore, we will try to utilize multi-FPGA architecture to improve the processing performance in the future.

\end{document}